\documentclass[a4paper,12pt]{article}

\usepackage{amsmath}
\usepackage[psamsfonts]{amssymb}
\usepackage{rsfs}
\usepackage{bm}

\usepackage{cite}
\usepackage[dvips]{graphicx}
\usepackage{color}

\makeatletter
\@addtoreset{equation}{section}
\makeatother

\begin{document} 

\begin{titlepage}

\baselineskip 10pt
\hrule 
\vskip 5pt
\leftline{}
\leftline{Chiba Univ. Preprint
          \hfill   \small \hbox{\bf CHIBA-EP-160}}
\leftline{\hfill   \small \hbox{hep-th/0609166}}
\leftline{\hfill   \small \hbox{November 2006}}
\vskip 5pt
\baselineskip 14pt
\hrule 
\vskip 1.0cm
\centerline{\Large\bf 
Gauge-invariant gluon mass, 
} 
\vskip 0.3cm
\centerline{\Large\bf  
infrared Abelian dominance
}
\vskip 0.3cm
\centerline{\Large\bf  
and 
stability of magnetic vacuum
}
\vskip 0.3cm
\centerline{\large\bf  
}

\vskip 0.5cm

\centerline{{\bf 
Kei-Ichi Kondo$^{\dagger, {1}}$
}}  
\vskip 0.5cm
\centerline{\it
${}^{\dagger}$Department of Physics, Faculty of Science, 
Chiba University, Chiba 263-8522, Japan
}
\vskip 0.3cm
\vskip 1cm

\begin{abstract}
We give an argument for deriving analytically the infrared ``Abelian'' dominance in a gauge invariant way for the Wilson loop average in SU(2) Yang--Mills theory.  In other words, we propose a possible mechanism  for  realizing the dynamical Abelian projection in the SU(2) gauge-invariant manner without breaking color symmetry. This supports validity of the dual superconductivity picture for quark confinement.  We also discuss the stability of the vacuum with magnetic condensation as a by-product of this result.

\end{abstract}

Key words:   quark confinement, Yang-Mills theory, Abelian dominance, gluon mass, magnetic monopole, monopole condensation,    
 
\vskip 0.5cm

PACS: 12.38.Aw, 12.38.Lg 
\hrule  
\vskip 0.1cm
${}^1$ 
  E-mail:  {\tt kondok@faculty.chiba-u.jp}

\par 
\par\noindent


\vskip 0.5cm

\newpage
\pagenumbering{roman}




\end{titlepage}


\pagenumbering{arabic}

\baselineskip 14pt
\section{Introduction}

Quark confinement is still an unsolved and challenging problem in theoretical particle physics, as is well known.  Though the approach to this problem is not unique, we have a promising scenario for explaining quark confinement, the so-called  dual superconductivity picture \cite{dualsuper} for the vacuum of the non-Abelian gauge theory \cite{YM54}.
This scenario proposed long ago is intuitively quite appealing.
Indeed, the relevant data supporting the validity of this picture have been accumulated by numerical simulations especially since 1990 and some of the theoretical predictions  \cite{tHooft81,EI82} have been confirmed by these investigations: infrared Abelian dominance \cite{SY90}, magnetic monopole dominance \cite{SNW94} and non-vanishing  off-diagonal gluon mass \cite{AS99} in the Maximal Abelian gauge \cite{KLSW87}, which are the most characteristic features for the dual superconductivity. 
In spite of these facts, the theoretical justification is not yet reached to a satisfactory level.
In this paper, we demonstrate analytically the infrared Abelian dominance in the Wilson loop average for a large Wilson loop of the SU(2) Yang-Mills theory \cite{YM54}.

For this purpose, we need to answer how to define and extract ``Abelian'' part $\mathbb{V}_\mu$ from the original non-Abelian gauge field $\mathscr{A}_\mu$, which is responsible for the area decay law of the Wilson loop average. This must be done without spoiling gauge invariance.  The conventional Abelian projection \cite{tHooft81} is too naive to realize this requirement.  
In sec. 2 and 3, we achieve this goal by using a non-Abelian Stokes theorem and a non-linear change of variables (NLCV). 

At the same time, we must answer why the ``remaining'' part $\mathbb{X}_\mu$ $(\mathbb{X}_\mu=\mathscr{A}_\mu-\mathbb{V}_\mu)$ in the non-Abelian gauge field $\mathscr{A}_\mu$ decouple in the low-energy (or long-distance) regime.  
To answer this question, we argue in sec. 4 and 5 that the remaining part $\mathbb{X}_\mu$ acquires the mass $M_X$ to be decoupled in the low-energy region.
The fundamental mechanism for the $\mathbb{X}_\mu$-mass generation is that the gauge-invariant composite operator $\mathbb{X}_\mu^2$ develops a non-vanishing vacuum expectation value $\left< \mathbb{X}_\mu^2 \right>$, in other words, the gauge-invariant dimension--two condensation takes place, i.e., $\left< \mathbb{X}_\mu^2 \right>\ne 0$. 
In fact, it was recently proposed in \cite{GSZ01} and \cite{Kondo01,Kondo03} that dimension--two vacuum condensations composed of gluon field are relevant to the realization of quark confinement and the existence of mass gap in Yang-Mills theory.
We also discuss some implications of dimension--two condensate $\left< \mathbb{X}_\mu^2 \right>$ for the low-energy description of Yang-Mills theory. 

Moreover, in sec. 5 we point out that the existence of such a condensation stabilizes the vacuum of the Savvidy type \cite{Savvidy77} with the magnetic condensation by eliminating a tachyon mode causing the Nielsen--Olesen instability \cite{NO78}.
This is a by-product of the above result.  
The stability of the magnetic vacuum is desirable for the magnetic monopole dominance. 
Thus the non-perturbative Yang-Mills vacuum is characterized by two vacuum condensations, i.e., the condensation $\left< \mathbb{X}_\mu^2 \right>\ne 0$ and the magnetic condensation $\left< H \right>\ne 0$, both of which realize the vacuum energy lower than that of the perturbative vacuum. 


\section{Non-Abelian Stokes theorem for the Wilson loop operator and  introduction of color  field}



For the non-Abelian gauge potential, $\mathscr{A}_\mu(x)=\mathscr{A}_\mu^A(x) T^A$, the Wilson loop operator $W_{\mathscr{A}}(C)$ for a closed loop $C$  is defined by
\begin{align}
 W_{\mathscr{A}}(C) 
 := \mathscr{N}^{-1} {\rm tr} \left[ \mathscr{P} \exp \left\{ ig \oint_{C} dx^\mu \mathscr{A}_\mu(x) \right\} \right] ,
\end{align} 
where $\mathscr{P}$ denotes the path-ordered product and the normalization factor $\mathscr{N}$ is equal to the dimension of the representation $R$, to which the probe of the Wilson loop belongs, i.e., $\mathcal{N}={\rm dim}({\bf 1}_R)={\rm tr}({\bf 1}_R)$.
Then the Wilson loop average $W(C):=\left< W_{\mathscr{A}}(C) \right>_{\rm YM}$, i.e., the vacuum  expectation value of the Wilson loop operator, is given by the   functional integration:
\begin{align}
W(C) := \left< W_{\mathscr{A}}(C) \right>_{\rm YM} 
=& Z_{\rm YM}^{-1} \int \mathcal{D}\mathscr{A}_\mu \exp \left( iS_{\rm YM} \right) W_{\mathscr{A}}(C) 
\nonumber\\
=& \frac{\int \mathcal{D}\mathscr{A}_\mu \exp \left( iS_{\rm YM} \right) W_{\mathscr{A}}(C) }{\int \mathcal{D}\mathscr{A}_\mu \exp \left( iS_{\rm YM} \right)},
\end{align} 
where $S_{\rm YM}$ is the Yang-Mills action.  
This expression is rather formal. For this to be a precise definition, we must specify the gauge fixing procedure to give a well-defined functional integration measure $\mathcal{D}\mathscr{A}_\mu$, as will be discussed later.

The Wilson loop operator is rewritten into a surface-integral form which is called the Non-Abelian Stokes theorem (NAST).  
We adopt in this paper the Diakonov--Petrov version \cite{DP89} of NAST
\footnote{
We adopt the coherent state representation for a derivation of the Diakonov--Petrov NAST as given in sec.III of \cite{KondoIV} for SU(2) and in \cite{KT99} for $SU(N), N \ge 3$. See also \cite{HU99} for more information. 
It was claimed that some care must be taken in using this version of NAST \cite{FITZ00}.
} 
which does not include neither the path ordering along the loop nor the surface ordering in sharp contrast to the other versions \cite{NAST} of NAST, at the price of an additional integration over  all gauge transformations of the given non-Abelian background field.  

The Diakonov--Petrov NAST is nothing but a path integral representation of the Wilson loop operator.  This representation is obtained according to the usual procedure of obtaining the path-integral representation: i) partitioning the closed loop $C$ into $N$ infinitesimal segments, ii) inserting the complete set at each partition point, iii) taking the limit $N \rightarrow \infty$ appropriately.  
As the complete set to be inserted, we use the coherent state which is described by introducing an auxiliary vector field $\bm{n}(x)$. 
The vector field $\bm{n}(x)$ is hereafter called color field  in relation to the Yang-Mills theory by the reason to be clarified later.

In what follows, we use a notation
$ F \cdot G  :=   F^A G^A = 2{\rm tr}(FG)$,  
 $F^2 := F \cdot F $,  
and
$
  (F \times G)^A := \epsilon^{ABC} F^B G^C =-2i {\rm tr}(T^A [F,G])
$ 
with
the normalization for the Hermitian generators $T^A$ of the Lie algebra $\mathscr{G}$ of the gauge group $G$: ${\rm tr}(T^A T^B)= \frac12 \delta^{AB}$.
For the gauge group $G=SU(2)$, 
the color field $\bm{n}(x)$ is the unit vector field with three components, i.e., 
\begin{equation}
 \bm{n}(x)=(n_1(x),n_2(x),n_3(x)) , \quad \bm{n}(x) \cdot \bm{n}(x) := n^A(x) n^A(x) =1 ,
\end{equation}
and  the path-integral representation reads
\begin{align}
 W_{\mathscr{A}}(C) 
 = \int d\mu_{C}[\bm{n}] \exp \left\{ i g J  \oint_{C} dx^\mu \{  {\rm tr}(\sigma_3 U \mathscr{A}_\mu U^\dagger) + ig^{-1} {\rm tr}(\sigma_3 U \partial_\mu U^\dagger)  \}  \right\}  ,
\end{align} 
where $J=\frac12, 1, \frac32, 2, \cdots$ is the index, say ``spin'', characterizing the representation $R$, to which the probe of the Wilson loop belongs, 
and $d\mu_{C}[\bm{n}]$ is the product measure of the normalized  invariant Haar measure $d\mu[\bm{n}(x)]$  on $SU(2)/U(1) \cong S^2$ at a spacetime point $x$: 
\begin{align}
 d\mu_{C}[\bm{n}]  := \prod_{x \in C} d\mu[\bm{n}(x)] ,
 \quad d\mu[\bm{n}(x)] = d^3\bm{n}(x) \delta(\bm{n}(x) \cdot \bm{n}(x)-1) .
\end{align}  
Here the unit vector field $\bm{n}(x)$ is defined through an SU(2) matrix field $U(x)$ by 
\begin{align}
 \hat{\bm{n}}(x) := n^A(x) \sigma_A  = U^\dagger(x) \sigma_3 U(x) \quad (A=1,2,3) ,
\end{align} 
with Pauli matrices $\sigma_A  (A=1,2,3)$.
Then, by using the Euler--angle representation
\begin{equation}
U(x)=e^{i\alpha(x)\sigma_3/2} e^{i\beta(x)\sigma_2/2} e^{i\gamma(x)\sigma_3/2} ,
\end{equation}
the color vector field $\bm{n}(x)$ is expressed by two Euler--angle fields $\alpha(x)$ and $\beta(x)$:  
\begin{equation}
 \bm{n}(x)=(n_A(x))_{A=1,2,3} =(\sin \beta(x) \cos \alpha(x), \sin \beta(x) \sin \alpha(x),  \cos \beta(x))  ,
\end{equation}
and an explicit form of the Haar measure is given by
\begin{equation}
d\mu[\bm{n}(x)]=\frac{2J+1}{4\pi} \sin \beta(x) d\beta(x) d\alpha(x) .
\end{equation}
Since the argument of the exponential is Abelian, we can use the ordinary Stokes theorem to rewrite the line integral to the surface integral:
\begin{align}
 W_{\mathscr{A}}(C) 
 = \int d\mu_{S}[\bm{n}] \exp \left\{ i g \frac{J}{2}  \int_{S:\partial S=C} d^2S^{\mu\nu} G_{\mu\nu} \right\}  ,
\end{align} 
where $S$ is an arbitrary surface spanned on the loop $C$, 
the antisymmetric tensor $G_{\mu\nu}$ is the curvature defined by
\begin{align}
 G_{\mu\nu}(x)
 =&  \partial_\mu [\bm{n}(x) \cdot \mathscr{A}_\nu(x)] - \partial_\nu [\bm{n}(x) \cdot \mathscr{A}_\mu(x)] - g^{-1} \bm{n}(x) \cdot (\partial_\mu \bm{n}(x) \times \partial_\nu \bm{n}(x)) ,
\label{Gmunu}
\end{align}
and $d\mu_{S}[\bm{n}]$ is the product measure over the surface $S$:
\begin{align}
 d\mu_{S}[\bm{n}]  := \prod_{x \in S} d\mu[\bm{n}(x)] .
\end{align}   
See e.g., \cite{KondoIV,KondoII} for details of the derivation.


\section{Non-linear change of variables for gluon fields}



\begin{figure}[htbp]
\begin{center}
\includegraphics[height=6.5cm]{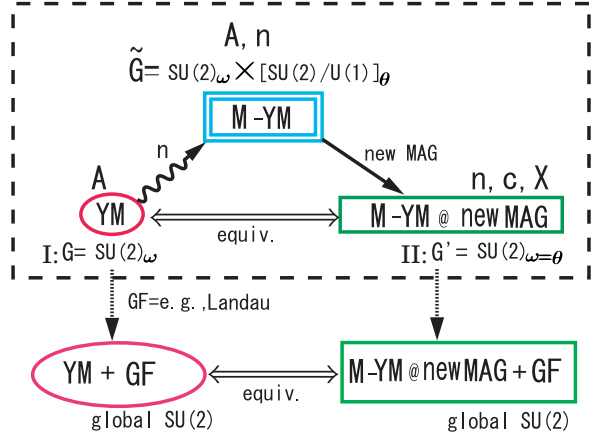}
\caption{\small 
The relationship between the original Yang-Mills (YM) theory and the master Yang--Mills (M-YM) theory. The master Yang--Mills theory has a larger (local and global) gauge group $\tilde{G}$ than the original gauge group $G$ of the original Yang-Mills theory and becomes equivalent to the original Yang-Mills theory after a constraint (new MAG) is imposed. The resulting gauge theory is denoted by M-YM at new MAG with the gauge group $G'$.
}
\label{fig:master-Yang-Mills}
\end{center}
\end{figure}


Since the loop $C$ can have arbitrary shape and arbitrary location in spacetime, the surface $S$ spanned by the loop $C$ may sweep the whole spacetime. 
In view of this, the color  field $\bm{n}(x)$ must be introduced over all the spacetime points.  
This is easily achieved by inserting the unity,
\begin{equation}
1=\int \mathcal{D}\mu[\bm{n}]
\equiv \int \mathcal{D}\bm{n}  \delta(\bm n\cdot\bm n-1)
:= \prod_{x \in \mathbb{R}^D} \int [d\bm{n}(x)]  \delta(\bm n(x) \cdot \bm n(x)-1) ,
\end{equation}
into the functional integration. 
Therefore, we arrive at the expression:
\begin{align}
W(C)  
=& \tilde{Z}_{\rm YM}^{-1} \int \mathcal{D}\mu[\bm{n}] 
 \int \mathcal{D}\mathscr{A}_\mu \exp \left( iS_{\rm YM} \right) \tilde{W}_{\mathscr{A}}(C) 
\nonumber\\
=& \frac{\int \mathcal{D}\mu[\bm{n}] \int \mathcal{D}\mathscr{A}_\mu \exp \left( iS_{\rm YM} \right) \tilde{W}_{\mathscr{A}}(C) }{\int \mathcal{D}\mu[\bm{n}] \int \mathcal{D}\mathscr{A}_\mu \exp \left( iS_{\rm YM} \right)},
\label{W1}
\end{align} 
where we have introduced the {\it reduced} Wilson loop operator $\tilde{W}_{\mathscr{A}}(C)$ defined by 
\begin{align}
 \tilde{W}_{\mathscr{A}}(C) 
 =  \exp \left\{ i g \frac{J}{2}  \int_{S:\partial S=C} d^2S^{\mu\nu} G_{\mu\nu} \right\}  ,
 \label{reducedW}
\end{align} 
and 
 the new partition function $\tilde{Z}_{{\rm YM}}$ has been defined by inserting the unity: 
$1=\int   \mathcal{D}\mu[\bm{n}]$ into the original Yang-Mills partition function:
\begin{align}
 \tilde{Z}_{{\rm YM}}  =\int \mathcal{D}\mu[\bm{n}] 
\int \mathcal{D}\mathscr{A}_\mu \exp (iS_{{\rm YM}}[\mathscr{A}]) . 
\label{Z}
\end{align}
At this stage, the color field $\bm{n}(x)$ is regarded as an auxiliary field introduced into the Yang-Mills theory in addition to the gauge field $\mathscr{A}_\mu(x)$.   We call this modified theory the  {\it master Yang-Mills theory}, which is written in terms of $\mathscr{A}_\mu(x)$ and $\bm{n}(x)$.  Thus we regard the vacuum expectation value of the Wilson loop in Yang-Mills theory as that of the reduced Wilson loop in the master Yang-Mills theory.  
The master Yang-Mills theory has more independent degrees of freedom than the original Yang-Mills theory.
For a while, we put this issue aside, until we will discuss how to reduce the master Yang-Mills theory to the original Yang-Mills theory shortly after introducing the non-linear change of variables.  
See Fig.~\ref{fig:master-Yang-Mills}.

We proceed to perform the (non-linear) change of variables of the original gauge field $\mathscr{A}_\mu$ by making use of the color field $\bm{n}(x)$. 
This will help us to clarify which variables are responsible for the area law of the Wilson loop average. 
Given a color field $\bm{n}(x)$, the Yang-Mills gauge field $\mathscr{A}_\mu(x)$ can be cast into the equivalent form: 
\begin{align}
 \mathscr{A}_\mu 
 =& (\bm{n} \cdot \mathscr{A}_\mu)\bm{n} + \mathscr{A}_\mu - (\bm{n} \cdot \mathscr{A}_\mu) \bm{n}
 \nonumber\\
 =& (\bm{n} \cdot \mathscr{A}_\mu)\bm{n} + (\bm{n}\cdot\bm{n}) \mathscr{A}_\mu - (\bm{n} \cdot \mathscr{A}_\mu) \bm{n}
 \nonumber\\
 =& (\bm{n} \cdot \mathscr{A}_\mu)\bm{n} + \bm{n} \times (\mathscr{A}_\mu \times \bm{n})
 \nonumber\\
 =& (\bm{n} \cdot \mathscr{A}_\mu)\bm{n} - g^{-1}\bm{n} \times \partial_\mu \bm{n} + g^{-1}\bm{n} \times (\partial_\mu \bm{n} + g\mathscr{A}_\mu \times \bm{n}) 
 \nonumber\\
 =& (\bm{n} \cdot \mathscr{A}_\mu)\bm{n} + g^{-1} \partial_\mu \bm{n} \times \bm{n} + g^{-1}\bm{n} \times D_\mu[\mathscr{A}]\bm{n}  ,
\label{CFNform}
\end{align}
where we have used only the relation $\bm{n}(x) \cdot \bm{n}(x)=1$ in the second equality
and introduced the covariant derivative in the last step:
\begin{align}
  D_\mu[\mathscr{A}] \bm{n}(x) 
  := \partial_\mu \bm{n}(x) +  g\mathscr{A}_\mu(x) \times \bm{n}(x) .
\label{XnA}
\end{align}
Thus the Yang-Mills gauge field $\mathscr{A}_\mu(x)$ is decomposed as 
\begin{equation}
\mathscr{A}_\mu(x)
 =c_\mu(x) \bm{n}(x)
  +g^{-1}\partial_\mu \bm{n}(x)\times \bm{n}(x)
  +\mathbb X_\mu(x) ,
\label{CFN}
\end{equation}
where we have used the identification:
\begin{align}
c_\mu(x)
 &={\bm n}(x)\cdot\mathscr{A}_\mu(x), 
\quad 
\mathbb X_\mu(x)
  =g^{-1}{\bm n}(x)\times D_\mu[\mathscr{A}]{\bm n}(x) .
\label{def:X}
\end{align}
The first term on the right-hand side of (\ref{CFN}) is denoted by 
$
  \mathbb{C}_\mu(x) := c_\mu(x){\bm n}(x) .
$
$\mathbb{C}_\mu(x)$ is parallel to $\bm{n}(x)$ and is called the restricted potential.
The second term is denoted by
$
  \mathbb{B}_\mu(x) := g^{-1}\partial_\mu{\bm n}(x)\times{\bm n}(x) .
$
$\mathbb{B}_\mu(x)$ is perpendicular to $\bm{n}(x)$ and is called the magnetic potential. 
For later convenience, we define $\mathbb{V}_\mu(x)$ by 
$\mathbb{V}_\mu(x):= \mathbb{C}_\mu(x) + \mathbb{B}_\mu(x)$:
\begin{align}
    \mathbb{V}_\mu(x)  
:= c_\mu(x) \bm{n}(x)
  +g^{-1}\partial_\mu \bm{n}(x)\times \bm{n}(x) .
\label{V}
\end{align}
As a way of specifying the separation of variables:
\begin{equation}
  \mathscr{A}_\mu(x) = \mathbb{V}_\mu(x) + \mathbb{X}_\mu(x) ,
\end{equation}  
the color field $\bm{n}(x)$ is required to be a covariant constant in the background field $\mathbb{V}(x)$:
\begin{align}
  D_\mu[\mathbb{V}] \bm{n}(x) := \partial_\mu \bm{n}(x) + g\mathbb{V}_\mu(x) \times \bm{n}(x)  = 0 .
  \label{D[V]n=0}
\end{align}
In fact, solving this equation for $\mathbb{V}_\mu(x)$ leads to (\ref{V}).
On the other hand, the remaining variable, i.e., the covariant potential  $\mathbb{X}_\mu(x)$ is required to be perpendicular to $\bm{n}(x)$:
\begin{align}
  \bm{n}(x) \cdot \mathbb{X}_\mu(x) = 0 .
  \label{nX=0}
\end{align}
This decomposition (\ref{CFN}) was once called the Cho-Faddeev-Niemi-Shabanov decompositions \cite{DG79,Cho80,FN98,Shabanov99} in the literatures. 
It is regarded as a non-linear change of variables (NLCV) for the original Yang-Mills field variables.

For our purposes, it is a remarkable fact that the curvature tensor $\mathscr{F}_{\mu\nu}[\mathbb{V}]$ obtained from the connection $\mathbb{V}_\mu$
is parallel to $\bm{n}$ and its magnitude $G_{\mu\nu}$   coincides exactly with the curvature tensor $G_{\mu\nu}$ appearing in the Wilson loop operator (\ref{Gmunu}) by way of the NAST: 
\begin{align}
 \mathscr{F}_{\mu\nu}[\mathbb{V}](x) 
:=& \partial_\mu \mathbb{V}_\nu(x)  - \partial_\nu \mathbb{V}_\mu(x)   + g \mathbb{V}_\mu(x)  \times \mathbb{V}_\nu(x) 
\nonumber\\
:=&  \bm{n}(x) G_{\mu\nu}(x) 
=\mathbb{G}_{\mu\nu}(x), \quad 
 \nonumber\\
 G_{\mu\nu}(x)  =&   \partial_\mu c_\nu(x) - \partial_\nu c_\mu(x) - g^{-1} \bm{n}(x) \cdot (\partial_\mu \bm{n}(x) \times \partial_\nu \bm{n}(x)) .
\end{align}
Therefore, we have succeeded to separate the original variables  
\begin{equation}
 (\mathscr{A}_\mu(x),\bm{n}(x)) \rightarrow 
(c_\mu(x), \mathbb{X}_\mu(x), \bm{n}(x)) ,
\end{equation}
with the identification (\ref{def:X}) such that only $\bm{n}(x)$ and $c_\mu(x)$ in the combined form $\mathbb{V}_\mu$ are responsible for the Wilson loop average and that the remaining variable $\mathbb{X}_\mu(x)$ is redundant for calculating the Wilson loop average. 
  In other words, $\mathbb{V}_\mu$ can be identified with the ``Abelian'' part of $\mathscr{A}_\mu$, suggesting the ``Abelian'' dominance in the Wilson loop average. 
 This fact has been already pointed out in the paper \cite{Cho00}.
However, this fact alone is not sufficient to guarantee the infrared Abelian dominance, since the theory has  interactions between $\bm{n}, c_\mu$ and $X_\mu$. 
 In order to confirm the infrared Abelian dominance, we must show that the variable $\mathbb{X}_\mu(x)$ is actually irrelevant for calculating the Wilson loop average.  This follows if these degrees of freedom decouple at least in the low-energy or long-distance region corresponding to the large Wilson loop.  A possible mechanism is discussed in what follows.

The SU(2) Yang--Mills Lagrangian density for the gluon field $\mathscr{A}_\mu$, 
\begin{align}
  \mathscr{L}_{YM}[ \mathscr{A}] :=  -\frac{1}{4} (\mathscr{F}_{\mu\nu}[\mathscr{A}])^2
=&  -\frac{1}{4} (\partial_\mu \mathscr{A}_\nu - \partial_\nu \mathscr{A}_\mu + g \mathscr{A}_\mu \times \mathscr{A}_\nu )^2 ,
\end{align}
is rewritten in terms of the new variables $(\bm{n}, c_\mu, \mathbb{X}_\mu)$ into 
\begin{align}
  \tilde{\mathscr{L}}_{YM}[\bm{n}, c, \mathbb{X}] = - \frac{1}{4} (\mathbb{G}_{\mu\nu}  + g \mathbb{X}_\mu \times \mathbb{X}_\nu )^2
- \frac{1}{4} (D_\mu[\mathbb{V}] \mathbb{X}_\nu - D_\nu[\mathbb{V}] \mathbb{X}_\mu )^2 , 
\end{align}
where we have used a fact that 
$\mathbb{G}_{\mu\nu}$ and $g \mathbb{X}_\mu \times \mathbb{X}_\nu$ 
are   parallel to 
$\bm{n}$, and this is also the case for the sum
$\mathbb{G}_{\mu\nu} + g \mathbb{X}_\mu \times \mathbb{X}_\nu$,
while $D_\mu[\mathbb{V}] \mathbb{X}_\nu - D_\nu[\mathbb{V}] \mathbb{X}_\mu$
 is orthogonal to $\bm{n}$ (which follows from the fact $\bm{n} \cdot \mathbb{X}_\mu=0$). 

By collecting the terms in $\mathbb{X}_\mu$, the Lagrangian reads
\begin{align}
 \tilde{\mathscr{L}}_{\rm YM}[\bm{n}, c, \mathbb{X}] = - \frac{1}{4} G_{\mu\nu}^2 - \frac{1}{2} X_\mu^{A} W_{\mu\nu}^{AB} X_\nu^{B} - \frac{1}{4} (g \mathbb{X}_\mu \times \mathbb{X}_\nu)^2 ,
\label{Lag}
\end{align} 
where
\begin{align}
  W_{\mu\nu}^{AB} 
 :=&   
  - \delta_{\mu\nu}  (D_\rho[\mathbb{V}] D_\rho[\mathbb{V}])^{AB} 
+  2g \epsilon^{ABC} n^C  G_{\mu\nu}   
+ (D_\mu[\mathbb{V}] D_\nu[\mathbb{V}])^{AB} .
\label{defW}
\end{align}
In this derivation, we have used the relation:
$[D_\mu[\mathbb{V}] , D_\nu[\mathbb{V}]]^{AB}=-g\epsilon^{ABC}\mathbb{G}_{\mu\nu}^C$.

We return to the issue raised above: how to reduce the master Yang-Mills theory to the original Yang-Mills theory. 
If we treated the color field $\bm{n}(x)$ as the fundamental field in addition to the original gauge field $\mathscr{A}_\mu(x)$, the resulting theory, say, the master Yang-Mills theory, had more independent degrees of freedom than those in the original Yang-Mills theory.  In other words, the master Yang-Mills theory had the  gauge symmetry $\tilde{G}:=SU(2)_{\omega} \times [SU(2)/U(1)]_{\theta}$ larger than the original gauge symmetry $G=SU(2)_{\omega}$. Here the latter symmetry $[SU(2)/U(1)]_{\theta}$ is carried by the color field $\bm{n}(x)$ while the former $SU(2)_{\omega}$ by the gauge field $\mathscr{A}_\mu(x)$.  In order to obtain the equivalent theory to the original Yang-Mills theory, we must impose necessary and sufficient numbers of constraints which eliminate the extra degrees of freedom and restrict the larger gauge symmetry to the $SU(2)$ gauge symmetry.   A suitable procedure was given in \cite{KMS06} by giving an explicit form of the constraint, which we called the new Maximal Abelian gauge (nMAG), although this naming is somewhat misleading.  
The nMAG is performed by minimizing the functional $\int d^D x \frac12 g^2 \mathbb X_\mu^2$ with respect to the enlarged gauge transformations: 
\begin{align}
 0 = \delta_{\omega, \theta} \int d^D x \frac12 g^2 \mathbb X_\mu^2 
= \delta_{\omega, \theta} \int d^D x   (D_\mu[\mathscr{A}]{\bm n})^2 .
 \label{MAGcond}
\end{align}
This determines the color field $\bm{n}(x)$ as a functional of a given configuration of $\mathscr{A}_\mu(x)$. 
The local gauge transformation of $\mathbb X^2$ is calculated as  \cite{KMS06}
\begin{align}
\delta_{\omega, \theta}\frac12\mathbb X_\mu^2
  = g^{-1}
     (D_\mu[\mathscr{A}]{\bm n}) \cdot 
     \{D_\mu[\mathscr{A}](\bm\omega_\perp - \bm\theta_\perp) \times \bm n \} ,
\label{eq:dX^2}
\end{align}
and the average over the spacetime of  (\ref{eq:dX^2}) reads  
\begin{align}
\delta_{\omega, \theta}\int d^D x\frac12\mathbb X_\mu^2
  =- \int d^D x
     (\bm\omega_\perp-\bm\theta_\perp)\cdot
     D_\mu[\mathbb V]\mathbb X_\mu .
 \label{minX2}
\end{align}
Hence, imposing (\ref{MAGcond}) 
for arbitrary $\bm\omega_\perp \ne \bm\theta_\perp$ 
yields a constraint in the differential form:
\begin{equation}
 \bm{\chi}
 :=D_\mu[\mathbb V]\mathbb X_\mu
 \label{dMAG}
 \equiv0 .
\end{equation}
This constraint yields two conditions, since $\bm{n} \cdot D_\mu[\mathbb V]\mathbb X_\mu=0$.     Imposing nMAG to the master Yang-Mills  theory breaks the enlarged $\tilde{G}:=SU(2)_{\omega} \times [SU(2)/U(1)]_{\theta}$ gauge symmetry  down to the diagonal SU(2) gauge symmetry: $G'=SU(2)_{\omega'}=SU(2)_{\rm II}$, a subgroup of $\tilde{G}$ ($\bm\omega \equiv \bm\theta:=\bm{\omega'}$).  The respective new variable transforms under this gauge transformation, say the local gauge transformation II as follows \cite{KMS06}.

\underline{Local gauge transformation II} (the active or background gauge transformation):
\begin{subequations}
\begin{align}
  \delta_{\omega}' \bm{n}(x)  =& g \bm{n}(x) \times \bm{\omega'}(x)  ,
\\
 \delta_{\omega}' c_\mu(x) =&    \bm{n}(x) \cdot \partial_\mu \bm{\omega'}(x)   ,
\\
  \delta_{\omega}' \mathbb{X}_\mu(x) =&  g \mathbb{X}_\mu(x) \times \bm{\omega'}(x) ,
\\
\Longrightarrow  
 &  \delta_{\omega}' \mathbb{V}_\mu(x) =   D_\mu[\mathbb{V}] \bm{\omega'}(x)    , 
\\
& \delta_{\omega}' \mathscr{A}_\mu(x)  
=   D_\mu[\mathscr{A}] \bm{\omega'}(x)  .
\end{align}
\end{subequations}
Therefore, $\mathbb X_\mu^2$ is invariant under the local gauge transformation II:
\begin{align}
 \delta_\omega' \mathbb{X}_\mu^2(x) =   0 .
\end{align}

The gauge transformation for the field strength is calculated using this result. The curvature $\mathbb{G}_{\mu\nu}:=\mathscr{F}_{\mu\nu} [\mathbb{V}]$  is subject to the adjoint rotation 
\begin{align}
  \delta_\omega' \mathbb{G}_{\mu\nu}(x) 
=& g \mathbb{G}_{\mu\nu}(x) \times \bm{\omega}'(x) .
\end{align}
Hence, the squared field strength has the   SU(2)$_{\rm II}$ invariance 
\begin{align}
 \delta_\omega' \mathbb{G}_{\mu\nu}(x)^2 
=&  0 .
\end{align}
The inner product of $\mathbb{G}_{\mu\nu}$ with $\bm{n}$, i.e., the magnitude of $\mathbb{G}_{\mu\nu}$, is also   SU(2)$_{\rm II}$ invariant:
\begin{align}
 \delta_\omega'( \bm{n}(x) \cdot  \mathbb{G}_{\mu\nu}(x)) 
 & \equiv \delta_\omega' G_{\mu\nu}(x) 
=    0 .
\end{align}
Thus the  {\it Abelian} field strength $G_{\mu\nu}$ is invariant under the SU(2) gauge transformation II, in sharp contrast to the original field strength $\mathscr{F}_{\mu\nu}$ which transforms in the adjoint representation.

In the functional integral formulation, we must specify the integration measure for the new variables. 
We must take into account the constraints 
$\delta(\bm{n}(x) \cdot \bm{n}(x)-1)$ and
$\delta(\bm{n}(x) \cdot \mathbb{X}_\mu(x))$ 
in the integration measure $\mathcal{D}\bm{n}(x) \mathcal{D}c_\mu(x) \mathcal{D}\mathbb{X}_\mu(x)$,
\begin{equation}
d\mu[\bm{n}] \mathcal{D}\mathscr{A}_\mu(x)
= \mathcal{D}\bm{n}(x) \delta(\bm{n}(x) \cdot \bm{n}(x)-1) \mathcal{D}c_\mu(x) \mathcal{D}\mathbb{X}_\mu(x) \delta(\bm{n}(x) \cdot \mathbb{X}_\mu(x)) .
\end{equation}  
To  avoid  complications coming from  constraints in performing the integration, we rewrite the integration measure in terms of the independent variables and calculate the Jacobian associated to this change of variables. 
For this purpose, we introduce the ortho-normal basis $(\bm{n}_1(x), \bm{n}_2(x), \bm{n}_3(x))=(\bm{e}_1(x), \bm{e}_2(x), \bm{n}(x))$, i.e., 
$\bm{n}_j(x) \cdot \bm{n}_k(x) = \delta_{jk}$, 
$\bm{n}_j(x) \times \bm{n}_k(x) = \epsilon_{jk\ell}\bm{n}_\ell(x)$, $(j,k=1,2,3)$, or equivalently
\begin{align}
   \bm{e}_a(x) \cdot \bm{e}_b(x) =& \delta_{ab} , \quad 
   \bm{n}(x) \cdot \bm{e}_a(x)  = 0, \quad
   \bm{n}(x) \cdot \bm{n}(x)  = 1
 \nonumber\\
  \bm{e}_a(x) \times \bm{e}_b(x) =& \epsilon_{ab} \bm{n}(x) , \quad
 \bm{n}(x) \times \bm{e}_a(x) = \epsilon_{ab}  \bm{e}_b(x)  , 
 \quad (a,b=1,2).
\label{basis}
\end{align}
The gauge transformation II for the basis vector $\bm{n}_j(x)$ is given by
\begin{equation}
 \delta_{\omega}' \bm{n}_j(x)  = g \bm{n}_j(x) \times \bm{\omega'}(x) 
 \Leftrightarrow 
 \delta_{\omega}' \bm{e}_a(x)  = g \bm{e}_a(x) \times \bm{\omega'}(x) , \quad
 \delta_{\omega}' \bm{n}(x)  = g \bm{n}(x) \times \bm{\omega'}(x) .
\end{equation}

It is easy to show that the Jacobian is equal to one for the transformation from the original variables $n^A, \mathscr{A}_\mu^B$ to the new variables 
$n^A, c_\mu, X_\nu^b$  in this basis $(\bm{e}_1(x), \bm{e}_2(x), \bm{n}(x))$ where (see Appendix~\ref{app:Jacobian} for the derivation):
\begin{align}
 & \mathbb{X}_\mu(x)  = X_\mu^a(x) \bm{e}_a(x)   
  \leftrightarrow  X^A_\mu(x) = X_\mu^a(x) e_a^A(x) 
\nonumber\\
& \text{or} \quad 
 X^a_\mu(x) = \mathbb{X}_\mu(x) \cdot \bm{e}_a(x) 
= X_\mu^A(x) e_a^A(x)   \quad (A=1,2,3: a=1,2) , 
\end{align}
so that  the integration measure is written as 
\begin{equation}
d\mu[\bm{n}] \mathcal{D}\mathscr{A}_\mu(x)
= \mathcal{D}n^a(x)  \mathcal{D}c_\mu(x) \mathcal{D}X^a_\mu(x)  .
\end{equation}  

Thus, we have given a reformulation of Yang-Mills theory in term of new variables obtained by using the non-linear change of variables.   
In this reformulated Yang-Mills theory, the Wilson loop average is given by
\begin{align}
W(C)  
=& \tilde{Z}_{\rm YM}^{-1} \int \mathcal{D}n^a(x)  \mathcal{D}c_\mu(x) \mathcal{D}X^a_\mu(x) \exp \left( i \tilde{S}_{\rm YM}[\bm{n},c,\mathbb{X}] \right) \tilde{W}_{\mathscr{A}}(C) 
\nonumber\\
=& \frac{\int\mathcal{D}n^a(x)  \mathcal{D}c_\mu(x) \mathcal{D}X^a_\mu(x) \exp \left( i \tilde{S}_{\rm YM}[\bm{n},c,\mathbb{X}] \right) \tilde{W}_{\mathscr{A}}(C) }{\int \mathcal{D}n^a(x)  \mathcal{D}c_\mu(x) \mathcal{D}X^a_\mu(x) \exp \left( i \tilde{S}_{\rm YM}[\bm{n},c,\mathbb{X}] \right)},
\label{W2}
\end{align} 
where we have omitted the gauge-fixing term corresponding to  nMAG (regarded as the condition for the partial gauge fixing from $\tilde{G}$ to $G'$) and the associated the Faddeev--Popov ghost and antighost term, see \cite{KMS05} for details.

By making use of the gauge-invariant Abelian field strength $G^{\mu\nu}$,  we can define the SU(2) {\it gauge-invariant} monopole current  by
\begin{align}
  k^\mu(x) :=& \partial_\nu {}^*G^{\mu\nu}(x) 
=   (1/2) \epsilon^{\mu\nu\rho\sigma}\partial_{\nu}
              G_{\rho\sigma}(x) .
\label{latCFN-monop}
\end{align}
Then the magnetic charge is defined by the volume integral of $k^0$:
\begin{equation}
 q_m := \int_{V} d^{3}x k^0  .
\end{equation}
This is cast into the surface integral over the closed surface $S$ as the boundary  of the volume $V$, $S=\partial V$:
\begin{equation}
 q_m  
 = \oint_{S} d^2S_\ell  {}^*G^{0\ell}  = \oint_{S} d^2S_\ell \frac12  \epsilon^{\ell jk} G_{jk} 
 = \oint_{S} d\sigma_{jk} G_{jk} . 
 \label{mcharge}
\end{equation}
It is easy to show that the gauge-invariant magnetic charge $q_m$ defined in this manner satisfies a charge quantization condition of the Dirac type:
\begin{equation}
 q_m = \frac{4\pi}{g} n \quad (n=0, \pm 1, \pm 2, \cdots ).
 \label{qc}
\end{equation}

Lattice formulations based on new variables of SU(2) Yang-Mills theory were for the first time constructed in \cite{KKMSSI06,IKKMSS06} so that they  reduce in the (naive) continuum limit to the reformulated SU(2) Yang-Mills theory written in terms of new variables obtained through NLCV from the original gauge field \cite{KMS05}. 
Two lattice formulations, non-compact \cite{KKMSSI06} and compact \cite{IKKMSS06}, enable one to define the magnetic monopole in the {\it gauge invariant} way keeping color symmetry  in Yang-Mills theory on a lattice without introducing fundamental scalar fields.
This is a remarkable result, since the conventional approach of defining the magnetic monopole in Yang-Mills theory without fundamental scalar fields heavily relies on a specific choice of gauge fixing, the so-called the maximal Abelian gauge (MAG) \cite{KLSW87}.  The MAG breaks the local $SU(2)$ gauge symmetry into the maximal torus group $U(1)$ and simultaneously SU(2) color symmetry into U(1). Therefore, only U(1) local and global gauge symmetry remain  if MAG is imposed. 
In the non-compact formulation \cite{KKMSSI06}, however, the magnetic charge resulting from the magnetic monopole defined in this way is not guaranteed to be integer-valued.  
The compact formulation \cite{IKKMSS06} was constructed so that the magnetic monopole defined on a lattice by the similar way to (\ref{mcharge}) is integer-valued and satisfies the quantization condition (\ref{qc}).   In fact, these features were directly confirmed by numerical simulations \cite{IKKMSS06}.


\section{Dynamical mass generation for the gluon fields $\mathbb{X}_\mu$ through the vacuum condensation of mass dimension--two $\left< \mathbb{X}_\mu^2 \right>$
}


We can introduce the gauge-invariant mass term which is invariant under the local SU(2) gauge transformation II as pointed out in \cite{KMS06}:
\begin{equation}
\mathscr{L}_{\rm m}
= \frac{1}{2} M_X^2 \mathbb{X}_\mu^2 .
\label{mass1}
\end{equation}
This  gauge-invariant mass term is rewritten in terms of the original variables $\mathscr{A}_\mu$:
\begin{align}
\mathscr{L}_{\rm m}
 =& \frac{1}{2} M_X^2 (\mathscr{A}_\mu - \mathbb{V}_\mu)^2 
 = \frac{1}{2} M_X^2 (\mathscr{A}_\mu -c_\mu \bm{n}
  +g^{-1}\partial_\mu \bm{n}\times \bm{n} )^2 
  \nonumber\\
=& \frac{1}{2g^2} M_X^2 (D_\mu[\mathscr{A}] \bm{n})^2 ,
  \label{mass2}
\end{align}
under the understanding that the color field $\bm{n}$ is expressed in terms of the original gauge field $\mathscr{A}_\mu$ by solving the nMAG constraint. 
Therefore, $\mathbb{V}_\mu$ (or $c_\mu$ and $\bm{n}$) plays the similar role to the St\"uckelberg field to recover the local gauge symmetry.
Note that   $c_\mu$, $\bm{n}$ and $\mathbb{X}_\mu$ are treated as independent variables after the non-linear change of variables and the mass term is a polynomial in the new variable $\mathbb{X}_\mu$, although they might be non-local and non-linear composite operators of the original variables $\mathscr{A}_\mu$. 
 
The proposed mass term (\ref{mass1}) or (\ref{mass2}) for the gluon should be compared with the conventional gauge-invariant mass term of Kunimasa--Goto type \cite{KG67}:
\begin{equation}
\mathscr{L}_{\rm KG}
 =   M^2 {\rm tr} \{(\mathscr{A}_\mu - ig^{-1}U\partial_\mu U^\dagger)^2 \}
 =   M^2 {\rm tr} \{(UD_\mu[\mathscr{A}]U^\dagger)^2 \} ,
 \quad 
 U(x) = e^{-i\chi(x)/v} .
 \label{mass3}
\end{equation}
This mass term is non-polynomial in the St\"uckelberg field $\chi(x)$.  This fact makes the field theoretical treatment very difficult.

We proceed to argue that there occurs a novel vacuum condensation of mass dimension--two for the field $\mathbb{X}^\mu$, i.e., $\left< -\mathbb{X}_\mu^2 \right> \ne 0$, 
\footnote{
We adopt the Minkowski metric $g_{\mu\nu}={\rm diag}(1,-1,-1,-1)$.  After the Wick rotation to the Euclidean region, the Minkowski metric tensor $g_{\mu\nu}$ is replaced by 
$
 -\delta_{\mu\nu} ={\rm diag}(-1,-1,-1,-1)
$.  Therefore, we have
$
 -\mathbb{X}_\mu^2 \rightarrow (\mathbb{X}^E_\mu)^2 > 0
$.
}
and that the field $\mathbb{X}^\mu$ acquires the mass dynamically through this condensation.  A naive way to see this is to use the mean-field like argument or the Hartree--Fock approximation which leads to 
the gauge-invariant mass term for  $\mathbb{X}_\mu$ gluons:
\begin{align}
 & -\frac{1}{4}(g \mathbb{X}_\mu \times \mathbb{X}_\nu) \cdot (g \mathbb{X}^\mu \times \mathbb{X}^\nu) 
 \nonumber\\
\to & \frac{1}{2}g^2 \mathbb{X}^A_\mu \left[\left\langle -\mathbb{X}^2_\rho \right\rangle \delta^{AB} - \left\langle -\mathbb{X}^A_\rho \mathbb{X}^B_\rho \right\rangle \right] \mathbb{X}^{\mu B} 
=   \frac{1}{2} M_X^2 \mathbb{X}_\mu \cdot \mathbb{X}^\mu ,
\quad M_X^2 = \frac23 g^2 \left\langle -\mathbb{X}^2_\rho \right\rangle .
\label{mass-term}
\end{align}

Now we make this idea more precise. 
First, we decompose the field $\mathbb{X}_\mu$ into the background field $\underline{\mathbb{X}}^\mu$ and the quantum fluctuation field $\tilde{\mathbb{X}}^\mu$ around it, i.e., $\mathbb{X}^\mu=\underline{\mathbb{X}}^\mu + \tilde{\mathbb{X}}^\mu$.  Then, we expand the Yang-Mills action for the Lagrangian density (\ref{Lag})  in $\tilde{\mathbb{X}}^\mu$ around $\underline{\mathbb{X}}^\mu$.  
The fact that the classical background field $\underline{\mathbb{X}}^\mu$  satisfies the equation of motion, 
$
 \frac{\delta \tilde{S}_{\rm YM}[n,c,\mathbb{X}]}{\delta X_\mu^A} \Big|_{\mathbb{X}=\underline{\mathbb{X}}}  = 0 
$,
yields up to quadratic in $\tilde{\mathbb{X}}^\mu$:
\begin{align}
 \tilde{S}_{\rm YM}[n,c, \mathbb{X}] 
=& \tilde{S}_{\rm YM}[n,c,\underline{\mathbb{X}}]  
 - \frac12 \tilde{X}^A_\mu \tilde{X}^B_\nu K_{\mu\nu}^{AB}  + \mathcal{O}(\tilde{\mathbb{X}}^3) ,
\end{align}
where the quartic self-interaction term generates additional quadratic terms in $\underline{\mathbb{X}}_\mu$:
\begin{align}
 K_{\mu\nu}^{AB} :=& - \frac{\delta^2 \tilde{S}_{\rm YM}[n,c,\mathbb{X}]}{\delta X_\mu^A \delta X_\nu^B}\Big|_{\mathbb{X}=\underline{\mathbb{X}}}  
\nonumber\\
=& \frac12 (W_{\mu\nu}^{AB} +  W_{\nu\mu}^{BA}) 
\nonumber\\
& + g^2[g_{\mu\nu} \delta^{AB} (\mathbb{X}_\rho)^2 - g_{\mu\nu} X_\rho^A X_\rho^B - \delta^{AB} X_\mu^C X_\nu^C + 2X_\mu^A X_\nu^B - X_\mu^B X_\nu^A ]|_{\mathbb{X}=\underline{\mathbb{X}}}   ,
\end{align}
with%
\footnote{
$K_{\mu\nu}^{AB}$ should be always understood in the quadratic form, $\frac12 \tilde{X}^A_\mu \tilde{X}^B_\nu K_{\mu\nu}^{AB}$. 
Therefore, 
$\frac12 (W_{\mu\nu}^{AB} +  W_{\nu\mu}^{BA})$
is equal to $W_{\mu\nu}^{AB}$ under this understanding. 
}
 $W_{\mu\nu}^{AB}$ defined by (\ref{defW}).
Hereafter, we neglect the quantum fluctuation parts of the other fields  $\bm{n}(x)$ and $c^\mu(x)$ by identifying them entirely with low-energy slowly-varying modes: 
 $\bm{n}(x)=\bm{\underline{n}}(x)$ and $c^\mu(x)=\underline{c}^\mu(x)$.  
This approximation will be improved by including also the high-energy modes for $\bm{n}(x)$ and $c^\mu(x)$ in a subsequent paper \cite{KSMO06}, although such an analysis has been tried in somewhat different context in \cite{Gies01}. 

By expanding the field $\mathbb{X}_\mu(x)$ in the basis ($\bm{e}_1(x), \bm{e}_2(x)$) which is perpendicular to $\bm{n}(x)$,
\begin{equation}
 \tilde{\mathbb{X}}_\mu(x)  = \tilde{X}_\mu^a(x) \bm{e}_a(x) , \quad \text{or} \quad 
 \tilde{X}^A_\mu(x) = \tilde{X}_\mu^a(x) e_a^A(x) \quad (A=1,2,3: a=1,2) , 
\end{equation}
the quadratic form is rewritten in terms of the independent fields $X_\mu^a (a=1,2)$:
\begin{equation}
 \frac12 \tilde{X}_\mu^A K_{\mu\nu}^{AB} \tilde{X}_\nu^B = \frac12 \tilde{X}_\mu^a K_{\mu\nu}^{ab} \tilde{X}_\nu^b , 
\quad
 K_{\mu\nu}^{ab} := e_a^A e_b^B  K_{\mu\nu}^{AB} 
 .
\end{equation}

Including the additional parts coming from quartic self-interactions among $\mathbb{X}_\mu$ gluons is equivalent to modify the  two-point gluon Green function (full gluon propagator) $\mathscr{D}_{\mu\nu}^{ab}$ by taking into account the tadpole contribution $\Pi_{\mu\nu}^{ab}$ of the background field:
\begin{align}
 K_{\mu\nu}^{ab} = (\mathscr{D}_{\mu\nu}^{ab})^{-1} =&  (D_{\mu\nu}^{ab})^{-1} +  \Pi_{\mu\nu}^{ab}   , 
\\
 (D_{\mu\nu}^{ab})^{-1} =& Q_{\mu\nu}^{ab} , 
\quad 
 \Pi_{\mu\nu}^{ab} =   V_{\mu\nu\rho\sigma}^{abcd} \underline{X}_\rho^c  \underline{X}_\sigma^d  , 
\label{SDE}
\end{align}
where the $\mathbb{X}_\mu$-independent part $Q_{\mu\nu}^{ab}(x)$ is defined by
\footnote{
In defining $Q_{\mu\nu}^{ab}$, we have dropped the last term 
$(D_\mu[\mathbb{V}] D_\nu[\mathbb{V}])^{AB}$
 coming from  $W_{\mu\nu}^{AB}$.  
If we introduce a gauge-fixing parameter $\alpha$ for nMAG, we have the gauge-fixing term of the form: 
$\int d^Dx \frac{1}{2\alpha}(D^\nu[\mathbb{V}]\mathbb{X}_\nu)^2
= - \int d^Dx  \frac{1}{2\alpha}\mathbb{X}_\mu^A (D^\mu[\mathbb{V}] D^\nu[\mathbb{V}])^{AB}\mathbb{X}_\nu^B$ which cancels the contribution 
$\frac{1}{2} \mathbb{X}_\mu^A (D^\mu[\mathbb{V}] D^\nu[\mathbb{V}])^{AB}\mathbb{X}_\nu^B$
 for $\alpha=1$.   
Or, such a contribution vanishes by taking into account the nMAG condition  $D^\nu[\mathbb{V}]\mathbb{X}_\nu=0$, which corresponds to $\alpha=0$.  
Therefore, the following calculations should be understood to be performed exclusively in these cases. 
Moreover, it is worth remarking that  exact satisfaction of nMAG condition (\ref{MAGcond}) is realized only in the case $\alpha=0$. 
} 
\begin{align}
  Q_{\mu\nu}^{ab}(x) :=&  \frac12 (W_{\mu\nu}^{AB}(x) +  W_{\nu\mu}^{BA}(x))  e_a^A(x) e_b^B(x)   
  =  g_{\mu\nu}  R^{ab}(x) 
+   2g G_{\mu\nu}(x)\epsilon^{ab}  ,
\nonumber\\
 R^{ab}(x) :=& \delta^{ab}[-\partial^2   + g^2 G_\rho(x)^2 ]
 + g \epsilon^{ab} [\ \partial_\rho G^\rho(x) + 2  G^\rho(x) \partial_\rho ] ,
\label{QR} 
\end{align} 
and $V_{\mu\nu\alpha\beta}^{abcd}$ is the four-point vertex for off-diagonal gluons \cite{SIK03}
\begin{equation}
 V_{\mu\nu\rho\sigma}^{abcd}
=  g^2 [\epsilon^{ab} \epsilon^{cd} I_{\mu\nu,\rho\sigma} + \epsilon^{ac} \epsilon^{bd} I_{\mu\rho,\nu\sigma} + \epsilon^{ad} \epsilon^{bc} I_{\mu\sigma,\nu\rho}  ] ,
\quad
 I_{\mu\nu,\rho\sigma} := (g_{\mu\rho}g_{\nu\sigma} - g_{\mu\sigma}g_{\nu\rho})/2 .
\end{equation} 
In deriving (\ref{QR}), we have used (\ref{basis}) to obtain
\begin{equation}
 D_\mu[\mathbb{V}] \bm{e}_a(x) = gG_\mu(x) \epsilon_{ab} \bm{e}_b(x) ,\quad
\end{equation}
where $G_\mu$  is the SU(2) gauge-invariant {\it Abelian} gauge field defined by
\begin{equation}
 G_\mu(x) = c_\mu(x) + h_\mu(x) , \quad
 h_\mu(x) = g^{-1} \partial_\mu \bm{e}_1(x)  \cdot  \bm{e}_2(x) 
 = - g^{-1} \partial_\mu \bm{e}_2(x)  \cdot  \bm{e}_1(x) .
\end{equation}
These relations are shown to hold in Appendix~\ref{app:cov-der}.
Eq.(\ref{SDE}) is nothing but the Schwinger-Dyson equation for the full gluon propagator in this approximation. This is a gap equation similar to that of four-fermion model of the Nambu--Jona-Lasinio type where the fermion mass is dynamically generated in a self-consistent way.

Suppose that neither color symmetry nor Lorentz symmetry are spontaneously broken.  Then we can not orient the component $X_\mu^a$ to a specific direction and the vacuum condensate is to be caused isotopically in color and Lorentz indices:  
\begin{align}
 \Pi_{\mu\nu}^{ab} 
= V_{\mu\nu\rho\sigma}^{abcd} \left< \underline{X}_\rho^c \underline{X}_\sigma^d  \right>   
= V_{\mu\nu\rho\sigma}^{abcd} \frac18 \delta^{cd} g_{\rho\sigma} \left< \underline{X}_\lambda^e \underline{X}_\lambda^e  \right>  
= \frac38  g^2 g_{\mu\nu} \delta^{ab} \left< \mathbb{\underline{X}}_\rho^2 \right>  . 
\end{align}
Thus, we conclude that the existence of the vacuum condensation $\left< -\mathbb{\underline{X}}_\rho^2 \right>$ generates the mass term for the gluon field $\mathbb{\tilde{X}}_\mu$:
\begin{equation}
  - \frac12 \tilde{X}_\mu^a K_{\mu\nu}^{ab} \tilde{X}_\nu^b 
= - \frac12 \tilde{X}_\mu^a Q_{\mu\nu}^{ab} \tilde{X}_\nu^b  +  \frac12 M_X^2  \tilde{\mathbb{X}}_\mu \cdot \tilde{\mathbb{X}}^\mu , \quad  M_X^2 := \frac38 g^2 \left< - \mathbb{\underline{X}}_\rho^2 \right> ,
\end{equation}
where 
$
 \mathbb{X}_\mu^2 = X_\mu^A X_\mu^A = X_\mu^a X_\mu^a 
$.
Then the $\mathbb{X}_\mu$ gluon modes  decouple in the low-energy (or long-distance) region below the mass scale $M_X$.  
Consequently, the infrared ``Abelian'' dominance for the large Wilson loop average follows immediately from the fact that the Wilson loop operator is written in terms of $\mathbb{V}_\mu$ alone.

The numerical simulations on a lattice \cite{IKKMSS06} have demonstrated the infrared ``Abelian'' dominance and magnetic monopole dominance in the string tension within our compact lattice formulation, although such phenomena were found for the first time in the MAG \cite{SY90,SNW94}.
In fact, the string tension calculated from the magnetic part of the Wilson loop average according to (\ref{W2}), (\ref{reducedW}) and (\ref{Z}) in this formulation reproduces  90 $\sim$ 95 \% of the full string tension calculated from the original Wilson loop average in the conventional lattice formulation. 
More numerical simulations \cite{SIKKMS06} have shown that the remaining field $\mathbb{X}_\mu$ defined on a lattice acquires the mass  
$M_X \cong 1.2 {\rm GeV}$ which is obtained as the exponential decay rate of the two-point correlation function $\left< \mathbb{X}_\mu(x) \cdot \mathbb{X}_\mu(y) \right>_{\rm YM}$ measured on a lattice: 
$\left< \mathbb{X}_\mu(x) \cdot \mathbb{X}_\mu(y) \right>_{\rm YM} \sim |x-y|^{-\alpha}\exp(-M_X|x-y|)$. 
This value agrees with that of the off-diagonal gluon mass in the MAG \cite{AS99}.  On the other hand, the same analysis applied to the ``Abelian'' gluon $\mathbb{V}_\mu$   leads to the result $M_V \cong 0.6 {\rm GeV}$. This is consistent with the ``Abelian'' dominance.


\section{Effective potential and the vacuum condensation}



In order to study which type of the vacuum is realized in Yang-Mills theory, we need to calculate the effective potential.
In particular, we pay attention to see whether such a dimension--two vacuum condensation occurs or not.  
First, we integrate out the fluctuation field $\tilde{\mathbb{X}}^\mu$ in the functional integration.  This is easily done for $X_\mu^a$ field with a trivial Jacobian for the change of variables:
\begin{align}
 \int \mathcal{D}\tilde{X}_\mu^a \exp \left\{ -i \frac{1}{2} \tilde{X}_\mu^a K_{\mu\nu}^{ab} \tilde{X}_\nu^b  \right\} 
 = (\det K_{\mu\nu}^{ab})^{-1/2} 
 = \exp \left[-i \frac{1}{2i} \ln \det K_{\mu\nu}^{ab} \right] .
\end{align} 
Therefore, we are to calculate the effective potential 
\begin{equation}
 V(X^2,G)=  \frac14 G_{\mu\nu}^2 
+ \frac{g^2}{4} (\mathbb{X}_\mu \times \mathbb{X}_\nu)^2 
+ \frac{1}{2i} {\rm tr} \ln  K_{\mu\nu}^{ab}  - \frac{1}{i}{\rm tr}\ln R^{ab} ,
\label{Veff0}
\end{equation}
where the quartic term in $\mathbb{X}_\mu$ is decomposed into two gauge-invariant pieces:
\begin{align}
  \frac{g^2}{4} (\mathbb{X}_\mu \times \mathbb{X}_\nu) \cdot (\mathbb{X}^\mu \times \mathbb{X}^\nu)
  =& \frac{g^2}{4} (\mathbb{X}_\mu \cdot \mathbb{X}^\mu)^2 - \frac{g^2}{4} (\mathbb{X}_\mu \cdot \mathbb{X}_\nu)^2 .
  \label{ida0}
\end{align}
Hereafter the underline for $\mathbb{X}_\mu$ will be omitted for simplifying the notation.  The final term in (\ref{Veff0}) comes from the integration over the ghost and antighost fields which are necessary to implement the nMAG correctly according to the BRST method, see  \cite{KMS05} for details. 
It should be remarked that this effective potential is SU(2) gauge invariant, since the {\it Abelian} gauge field $G_\mu$ is SU(2) gauge-invariant as well as the Abelian field strength $G_{\mu\nu}$, see Appendix~\ref{app:transf}.  This is an advantage of our formulation, contrary to the conventional approach. 

For our purposes, we  examine the effective potential as a function of two vacuum condensates $\mathbb{X}_\mu^2$ and $G_{\mu\nu}$:
\begin{equation}
 V(X^2,G)=  \frac14 G_{\mu\nu}^2 
+ \frac{g^2}{4} (\mathbb{X}^2)^2
+ \frac{1}{2i} {\rm tr} \ln  K_{\mu\nu}^{ab}  - \frac{1}{i}{\rm tr}\ln R^{ab} ,
\end{equation}
where $K_{\mu\nu}^{ab}$ is shifted from $Q_{\mu\nu}^{ab}$ defined in (\ref{QR}) as
\begin{equation}
 K_{\mu\nu}^{ab} = Q_{\mu\nu}^{ab} + \frac38 g^2  \mathbb{X}^2 g_{\mu\nu} \delta^{ab} .
\end{equation}
In calculating the effective potential,  $G_{\mu\nu}$ and $X^2$ are assumed to be constants uniform in space and time.

In the limit of vanishing Abelian condensation $G_{\mu\nu}=0$, the effective potential  $V(X^2,0)$ is written in the closed form:
\begin{equation}
 V(\phi,0)=  \frac{16}{9} \frac{1}{g^2} \phi^2 + \frac{2}{(4\pi)^2}  \phi^2   \left( \ln \frac{\phi}{\mu^2} - \frac32  \right)  , 
\quad \phi := \frac38 g^2 \mathbb{X}^2  .
\end{equation}

This is calculated as follows. 
The dimensional regularization yields 
\begin{align}
 &  \ln \det \{   [-\partial^2 + \phi]  \}  
  = {\rm tr}\ln   \{   [-\partial^2 + \phi]  \}  
\nonumber\\
&=  - \frac{\Gamma(-D/2)}{(4\pi)^{D/2}} \phi^{D/2}
= -    \frac{\Gamma(-2+\epsilon)}{(4\pi)^{2-\epsilon}} \phi^{2-\epsilon}
  \nonumber\\
&= - \frac{1}{2} \frac{\phi^2}{(4\pi)^{2}} \left[ \epsilon^{-1} + \ln 4\pi - \gamma_E  - \ln \frac{\phi}{\mu^2} + \frac32 + \mathcal{O}(\epsilon) \right] ,
\end{align} 
where $\epsilon := (4-D)/2$ and $\gamma_E$ is the Euler constant. 
Then we obtain
\begin{equation}
 V(\phi,0)=  \frac{16}{9} \frac{1}{g^2} \phi^2 -  \frac{2\phi^2}{(4\pi)^{2}} \left[ \epsilon^{-1} + \ln 4\pi - \gamma_E  - \ln \frac{\phi}{\mu^2} + \frac32 + \mathcal{O}(\epsilon) \right] . 
\end{equation}

\begin{figure}[htbp]
\begin{center}
\includegraphics[height=5cm]{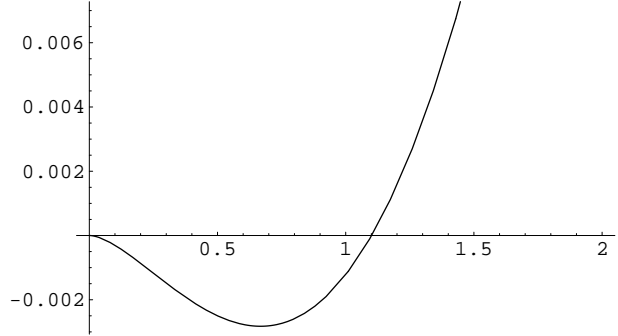}
\caption{\small 
The effective potential $V(\phi,0)$ where $M=1$ and $g=10$. 
}
\label{figure:effetive-potential}
\end{center}
\end{figure}

We introduce renormalization constants for $\phi$ and $g$ as
\begin{equation}
 \phi = Z_\phi^{1/2} \phi_R ,
\quad 
 g = Z_g g_R . 
\end{equation}
Then the effective potential reads 
\begin{equation}
 V(\phi,0)=  \frac{16}{9} \frac{1}{g_R^2} \phi_R^2 Z_g^{-2} Z_\phi -  \frac{2\phi_R^2 Z_\phi}{(4\pi)^{2}} \left[ \epsilon^{-1} + \ln 4\pi - \gamma_E  - \ln \frac{\phi_R  Z_\phi^{1/2}}{\mu^2} + \frac32  \right]  . 
\end{equation}
 From the observation, 
\begin{equation}
 \frac{1}{g^2} \phi^2 
= \frac{1}{g_R^2} \phi_R^2 Z_g^{-2} Z_\phi 
=  \frac{1}{g_R^2} \phi_R^2  -  \frac{1}{g_R^2} \phi_R^2 (1 - Z_g^{-2} Z_\phi )  ,
\end{equation}
and
\begin{align}
 &  \frac{\phi_R^2 Z_\phi}{(4\pi)^{2}} \left[ \epsilon^{-1} + \ln 4\pi - \gamma_E  - \ln \frac{\phi_R  Z_\phi^{1/2}}{\mu^2} + \frac32  \right] 
  \nonumber\\
=& \frac{\phi_R^2}{(4\pi)^{2}} \left[ \epsilon^{-1} + \ln 4\pi - \gamma_E  - \ln \frac{\phi_R  Z_\phi^{1/2}}{\mu^2}+ \frac32  \right]   
  \nonumber\\& 
+ \frac{\phi_R^2}{(4\pi)^{2}} (Z_\phi-1)\left[ \epsilon^{-1} + \ln 4\pi - \gamma_E  - \ln \frac{\phi_R  Z_\phi^{1/2}}{\mu^2}  + \frac32  \right]   
  \nonumber\\
=& \frac{\phi_R^2}{(4\pi)^{2}} \left[  - \ln \frac{\phi_R}{M^2} + \frac32    \right]
+ \frac{\phi_R^2}{(4\pi)^{2}} \left[ \epsilon^{-1} + \ln 4\pi - \gamma_E  -  \ln \frac{M^2}{\mu^2}  \right]   
  \nonumber\\& 
+ \frac{\phi_R^2}{(4\pi)^{2}} \left\{  (Z_\phi-1) \left[ \epsilon^{-1} + \ln 4\pi - \gamma_E  - \ln \frac{\phi_R  Z_\phi^{1/2}}{\mu^2} + \frac32  \right]  
-  \ln Z_\phi^{1/2} \right\}   ,
\end{align}
these renormalization constants are chosen so that the counter term 
$
 \frac{1}{g_R^2} \phi_R^2 (1 - Z_g^{-2} Z_\phi )
$
cancels the divergence in the bare effective potential, namely, to the lowest order of the coupling $g_R^2$, 
\begin{equation}
 \frac{16}{9} \frac{1}{g_R^2}   (1 - Z_g^{-2} Z_\phi ) 
+   \frac{2}{(4\pi)^{2}} \left[ \epsilon^{-1} + \ln 4\pi - \gamma_E  -  \ln \frac{M^2}{\mu^2}  \right] =0 ,
\label{renor}
\end{equation}
where we have used $Z_\phi, Z_g=1+\mathcal{O}(g^2)$.
Thus we obtain the renormalized effective potential:
\begin{equation}
 V_R(\phi_R,0)=  \frac{16}{9} \frac{1}{g_R^2} \phi_R^2 + \frac{2}{(4\pi)^2}  \phi_R^2   \left( \ln \frac{\phi_R}{M^2} - \frac32  \right)  ,
\end{equation}
which satisfies the renormalization condition:
$
 V''(\phi_R=M^2,0)=(32/9)/g^2 .
$
Indeed, the potential $V(\phi_R,0)$ has  a minimum at $\phi_R=\phi_R^0 \ne 0$  away from the origin: 
\begin{equation}
 \phi_R^0 = M^2 e^{1/2} \exp \left[ - \frac{(4\pi)^2}{(9/8)g_R^2(M)} \right] .
\end{equation}
See Fig.~\ref{figure:effetive-potential}.
This results shows that $\mathbb{X}_\mu^2$ condensates indeed. 
In order to establish this condensation, this one-loop result should be improved.


Operating the differential operator $\mu \frac{\partial }{\partial \mu}$ to (\ref{renor})  and 
defining the $\beta$-function $\beta(g_R)$ and the anomalous dimension $\gamma_\phi(g_R)$ as functions of $g_R$ by
\begin{align}
 \beta(g_R) = \mu \frac{\partial g_R}{\partial \mu} 
= - g_R  \mu \frac{\partial \ln Z_g}{\partial \mu} , \quad 
 \gamma_\phi(g_R) = \frac12  \mu \frac{\partial \ln Z_\phi}{\partial \mu}
= - \mu \frac{\partial \ln \phi_R}{\partial \mu} ,
\end{align}
we obtain a relationship between $\beta(g_R)$ and $\gamma_\phi(g_R)$ up to $\mathcal{O}(g^3)$:
\begin{align}
 \beta(g_R)+ g_R \gamma_\phi(g_R) = \frac{\frac98}{(4\pi)^2} g_R^3  .
\end{align}
This is consistent with the asymptotic freedom \cite{KondoI},
$
 \beta(g_R) = - \frac{b_0}{(4\pi)^2} g_R^3   
$
with
$
 b_0 = \frac{22}{3}
$, 
 provided that
\begin{align}
  \gamma_\phi(g_R) = g_R^{-1} \left[ \frac{\frac98}{(4\pi)^2} g_R^3 - \beta(g_R) \right] 
=   \frac{\frac98+b_0}{(4\pi)^2} g_R^2 .
\end{align}
In other words, the renormalized effective potential satisfies the renormlization group equation:
\begin{align}
  \left[ M \frac{\partial }{\partial M} + \beta(g_R) \frac{\partial }{\partial g_R} -  \gamma_\phi(g_R) \phi_R \frac{\partial }{\partial\phi_R} \right] V_R(\phi_R) = 0 .
\end{align}
Calculating the anomalous dimension $\gamma_\phi$ in consistent with this relation will be given elsewhere \cite{KSMO06}.

This result should be compared with the effective potential $V(\sigma)$ calculated in \cite{Dudal} for the gluon--ghost mixed composite operator of mass dimension--two, $\frac12 A_\mu^a A_\mu^a + i\alpha \bar{C}^a C^a$, which is shown \cite{Kondo01} to be on-shell BRST and anti-BRST invariant in the modified MA gauge \cite{KondoII}.  
However, the vacuum energy $E=V(\sigma_0)$ reached at the minimum depended strongly on the gauge-fixing parameter $\alpha$ of MA gauge.  This indicates that the vacuum realized at $\sigma_0$ does not corresponds to the true vacuum. 
Our effective potential $V(\phi,G)$ is guaranteed to be gauge invariant by construction.  Therefore, we can choose any value of gauge fixing parameter for nMAG.   
In particular, the choice $\alpha=1$ simplifies the calculations so that the final term in $W_{\mu\nu}^{ab}$ cancels with a gauge-fixing term of nMAG. This choice is also preferred from a fact that the mixed composite operator for $\alpha=1$ reduces to the gauge-invariant part of the gluon composite operator $\frac12 A_\mu^a A_\mu^a$, as demonstrated in \cite{Kondo03}. 
The choice $\alpha=0$ is the best for realizing the nMAG condition exactly.  
 The true vacuum should be obtained at which the potential $V(\phi,G)$ takes the minimum. Thus the total effective potential could be gauge parameter independent, if the contribution from the Abelian part is included, even in the MA gauge. 

In the lattice formulation, the effective potentials for various composite operators of mass dimension--two \cite{GSZ01,Kondo01,Kondo03} were calculated numerically on a lattice  \cite{KKMSS05}.   The result shows that the numerically obtained effective potential for $\mathbb{X}_\mu^2$ has a minimum away from the origin, suggesting the existence of non-vanishing vacuum condensates $\left< \mathbb{X}_\mu^2 \right> \ne 0$. 
However, it should be remarked that they are unrenormalized quantities.  We need to perform the non-perturbative renormalization to obtain the definite result for the existence of such a dimension--two condensate. Therefore, it is not yet confirmed whether they survive in the continuum limit.


\section{Stability of the magnetic condensation}


Finally, we examine the contribution from the Abelian part $G_{\mu\nu}=(\bm{E},\bm{H})$.  
We distinguish two cases:
(I) $\bm{E} \cdot \bm{H} \ne 0$, 
(II) $\bm{E} \cdot \bm{H} =0$, 
characterized by two Lorentz invariants defined by
\begin{align}
 \mathcal{F} :=& (\bm{E}^2-\bm{H}^2)/2 = - \frac12 G_{\mu\nu} G^{\mu\nu} = (a^2-b^2)/2 ,
 \\
 \mathcal{G} :=& \bm{E} \cdot \bm{H} = - \frac14 G_{\mu\nu}{}^*G^{\mu\nu} = ab ,
\end{align}
where
$
 a = \sqrt{\sqrt{\mathcal{F}^2+\mathcal{G}^2}+\mathcal{F}}, 
$
and
$
 b = \sqrt{\sqrt{\mathcal{F}^2+\mathcal{G}^2}-\mathcal{F}} .
$

(I) If $\bm{E} \cdot \bm{H} \ne 0$, i.e., $\mathcal{G}\ne 0$, it is possible to transform to a Lorentz frame in which $\bm{E}$ and $\bm{H}$ are parallel or anti-parallel depending on the signature of $\bm{E} \cdot \bm{H}$. 
We can choose the $z$ axis as the direction of the vector  without loss of generality: $\bm{E}=(0,0,E)$ and $\bm{H}=(0,0,H)$. 
The self-dual (or anti self-dual) case is a special case of (I):
$G_{\mu\nu}={}^*G_{\mu\nu}$ (or $G_{\mu\nu}=-{}^*G_{\mu\nu}$), i.e.,  $\bm{E}=(0,0,E)=\bm{H}=(0,0,H)$ (or  $\bm{E}=(0,0,E)=-\bm{H}=(0,0,-H)$).

(II) If $\bm{E} \cdot \bm{H} =0$, i.e., $\mathcal{G}=0$ in a Lorentz frame, $a=0$ or $b=0$;  $\bm{E}$ and $\bm{H}$ are also perpendicular in any other Lorentz frame, since $\bm{E} \cdot \bm{H}$ is a Lorentz invariant. 
If $\bm{E}^2>\bm{H}^2$, i.e., $\mathcal{F}>0$, then the situation is as for the purely electric field with a vacuum instability,  as is well known in QED.  On the other hand, if $\bm{H}^2>\bm{E}^2$, i.e., $\mathcal{F}<0$, then the system behaves like the case of a purely magnetic field.  If $\bm{E}^2=\bm{H}^2$, then we have a trivial case $\mathcal{F}=0$.

The trace of the logarithm of a matrix is calculated, once all the eigenvalues of the matrix  are known. 
In the pure magnetic case of (I), we can obtain the closed form of the effective potential, since the eigenvalues for $R^{ab} $ and $\tilde{Q}_{\mu\nu}^{ab}$  are exactly obtained to be capable of   summing up all the contributions coming from all the eigenvalues.  By using the same method as in \cite{Kondo04,Kondo05} and \cite{Flory83}, we obtain 
\begin{align}
& V(\phi,H)
 \nonumber\\
=&  \frac{1}{4g^2} (gH)^2 
 + \frac{16}{9} \frac{1}{g^2} \phi^2
 - \frac{1}{(4\pi)^2} \frac14 (gH)^2 \Big\{ 
 \nonumber\\
& 8 \Big[\zeta(-1,\frac{3+r}{2})+\zeta(-1,\frac{-1+r}{2})
+\zeta(-1,\frac{1+r}{2})-\zeta(-1,\frac{1}{2}) \Big] \left( \ln \frac{gH}{\mu^2}+c \right) 
 \nonumber\\
&+ 8\Big[\zeta^\prime(-1,\frac{3+r}{2})+\zeta^\prime(-1,\frac{-1+r}{2})+\zeta^\prime(-1,\frac{1+r}{2})-\zeta^\prime(-1,\frac{1}{2}) \Big] \Big\}   , \quad r := \frac{\phi}{gH} ,
\end{align}
where 
$ 
 c:=-1+\gamma_E + \ln 2
$
with an Euler constant $\gamma_E =0.5772\cdots$, 
and 
$\zeta(s,z)$ is called the generalized zeta function or Hurwitz $\zeta$-function defined by
\begin{equation}
 \zeta(s,z) := \sum_{n=0}^{\infty} (n+z)^{-s}, \quad 
 \Re(s) > 1, \quad z \ne 0, -1, -2, \cdots .
\end{equation}
The generalized zeta function  has an integral representation:
\begin{equation}
 \zeta(s,z) = \frac{1}{\Gamma(z)} \int_{0}^{\infty} dt \frac{e^{-zt} t^{s-1}}{1-e^{-t}} ,   \quad
 \Re(s) > 1, \quad \Re(z) > 0 ,
\end{equation}
and its prime denotes the differentiation with respect to the first variable:
$
 \zeta^\prime(s,z) :=\frac{d}{ds} \zeta(s,z)
$.
The zeta function is a special case of $z=0$ of the generalized zeta function:
$\zeta(s) = \zeta(s,0)$.
The generalized zeta function can be analytically continued in the complex $s$ plane to define an analytic function with a single simple pole at $s=1$. 

In the limit $r \rightarrow 0$, i.e., neglecting the effect of vacuum condensation for off-diagonal gluons $\mathbb{X}$, we recover the Nielsen--Olesen result \cite{NO78}:
\begin{align}
 V(0,H)
= \frac14 H^2 
+ \frac{1}{(4\pi)^2} \frac14 g^2 H^2 \Big\{ 
\frac{22}{3} \left( \ln \frac{gH}{\mu^2}+c' + 4\pi i \right) 
 \Big\}  ,
\end{align}
where 
$ 
 c':=c + \frac{24}{11}\zeta^\prime(-1,\frac32 ) 
= -1+\gamma_E + \ln 2 + \frac{24}{11}\zeta^\prime(-1,\frac32 ) 
= -0.94556 \cdots
$
with 
$\zeta^\prime(-1,\frac32 )=-0.817409\cdots$.
Here we have used
\begin{align}
 \zeta(-1,\frac32 ) =& - \frac{11}{24} = \zeta(-1, -\frac12 ) ,
 \\
  8 \left[\zeta(-1,\frac{3}{2})+\zeta(-1,\frac{-1}{2})\right] =& - \frac{22}{3} ,
 \\
 8 \left[\zeta^\prime(-1,\frac32 ) + \zeta^\prime(-1, \frac{-1}{2}) \right] 
=& 8[2\zeta^\prime(-1,\frac32 ) - i \frac{\pi}{2} ]
= 16  \zeta^\prime(-1,\frac32 ) - i 4\pi .
\end{align}
This potential $V(0,H)$ exhibits the instability due to the non-vanishing pure imaginary part, 
$\frac{i}{4\pi}\frac{11}{3}g^2H^2$.  This is the so-called Nielsen--Olesen instability \cite{NO78} to the magnetic vacuum of the Savvidy type \cite{Savvidy77}. 

The Nielsen--Olesen  instability survives as long as $r<1$. 
If $r \ge 1$, however,  the pure imaginary part vanishes and the effective potential $V(\phi,H)$ becomes real number.
This is because 
the four terms in front of $\ln gH/\mu^2$ can be simplified by using 
\begin{equation}
 \zeta(-1,z) = -\frac12 (z^2 - z + \frac16 ) ,
\end{equation}
as
$
 8 \Big[\zeta(-1,\frac{3+r}{2})+\zeta(-1,\frac{-1+r}{2})+\zeta(-1,\frac{1+r}{2})-\zeta(-1,\frac{1}{2}) \Big]
 = - \frac{22+9r^2}{3}
$
and the primed zeta function 
$\zeta^\prime(-1,z)$ is real for $z \ge 0$.
The last statement is checked by using the identity \cite{Dunne04}:
\begin{equation}
 \zeta^\prime(-1,z) = \zeta^\prime(-1) - \frac{z}{2} \ln (2\pi) - \frac{z}{2}(1-z) + \int_{0}^{z} dx \ln \Gamma(x) ,
\end{equation}
following from an integration of Binet's integral representation \cite{Erdelyi81,WW27} of $\ln \Gamma(x)$ 
and the Taylor expansion \cite{Erdelyi81,AS72,GR72} of $\ln \Gamma(x)$ 
\begin{equation}
 \ln \Gamma(x)  =  -\ln x - \gamma x + \sum_{n=2}^{\infty} \frac{(-1)^n}{n} \zeta(n) x^n  ,
\end{equation}
where
$\zeta^\prime(-1)=\zeta^\prime(-1,0)=-0.1654 \cdots$.

\begin{figure}[htbp]
\begin{center}
\includegraphics[height=4cm]{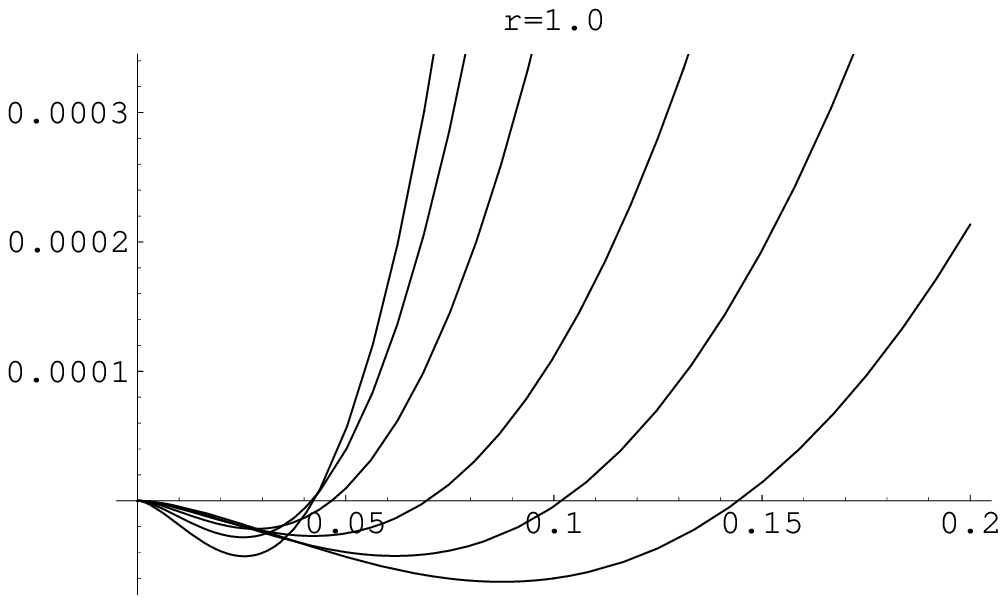}
\includegraphics[height=4cm]{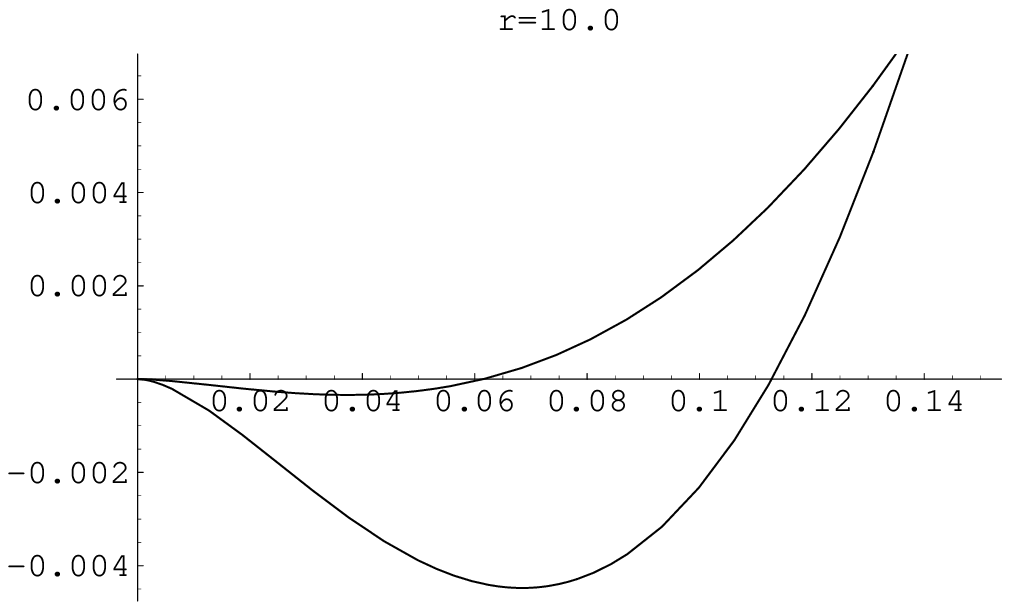}
\caption{\small 
The effective potential $V(\phi,H)$ vs. $gH$ with a fixed value of $r:=\phi/(gH)$. (Left panel) $r=1.0$,  $1.5$, $2.0$, $3.0$, $4.0$, $5.0$ from right to left.
(Right panel)
$r=10.0$, $20.0$.
}
\label{figure:effetive-potential2}
\end{center}
\end{figure}

The absolute minimum for $V(\phi,H)$ exists in the region $r > 1$, and the value of $r$ is determined as the point at which the potential takes the minimum. 
See the plot of $V(\phi,H)$ in Fig.~\ref{figure:effetive-potential2}.
Thus, the existence of dimension--two condensate $\left< \mathbb{X}^2 \right> \ne 0$ guarantees as a by-product the stability of the magnetic condensation $\left< H \right> \ne 0$ of the Savvidy type.   
In other words, the existence of dimension--two condensate $\left< \mathbb{X}^2 \right> \ne 0$ shifts  the gluon spectrum  upward and eliminates the  tachyonic mode causing the Nielsen--Olesen instability to recover the stability of the vacuum with magnetic condensation. 


\section{Conclusion and discussion}


In this paper, we have separated the original SU(2) gluon field variables $\mathscr{A}_\mu$ into  $\mathbb{V}_\mu$ and  $\mathbb{X}_\mu$ so that the variables $\mathbb{V}_\mu$ are responsible for quark confinement and the remaining variables $\mathbb{X}_\mu$ could decouple in the low-energy region.
  The former comes from a fact that a version of the non-Abelian Stokes theorem enables us to rewrite the non-Abelian Wilson loop operator entirely in terms of the SU(2) invariant {\it Abelian} field strength $G_{\mu\nu}$ defined from the variable $\mathbb{V}_\mu$.  
For the latter, we have argued that the gluon $\mathbb{X}_\mu$ acquires the gauge-invariant mass dynamically through the non-vanishing vacuum condensation of mass dimension--two $\left< \mathbb{X}_\mu^2 \right> \ne 0$.

We have given a first analytical calculation of the effective potential of the composite gluon operator of mass dimension--two $\mathbb{X}_\mu^2$ demonstrating the occurrence of the condensation as realized at the minimum located away from the origin.  See \cite{KKMSS05} for numerical simulations for the effective potential on a lattice.  Note that the composite operator $\mathbb{X}_\mu^2$ is gauge-invariant and the resulting mass term for $\mathbb{X}_\mu$ can be induced keeping the original SU(2) gauge invariance intact, contrary to the conventional wisdom.  Consequently, the decoupling of these degrees of freedom is characterized as a gauge-invariant low-energy phenomenon. In other words, this is a dynamical Abelian projection, suggesting the validity of the dual superconductor picture for quark confinement.  Thus the infrared ``Abelian'' dominance  immediately follows in the gauge invariant manner 

A next issue to be investigated is to derive analytically the area law for the Wilson loop average rewritten in term of new variables $\mathbb{V}_\mu$, since the area law with a string tension reproducing the full string tension was already shown numerically on a lattice \cite{IKKMSS06}, confirming the infrared ``Abelian'' dominance in the numerical way. 
The magnetic monopole is defined through $\mathbb{V}_\mu(x)$ namely $c_\mu(x)$ and $\bm{n}(x)$ in the gauge-invariant way even in the Yang-Mills theory without introducing any scalar field as a fundamental field, just as the 'tHooft and Polyakov monopole in the Georgi-Glashow model. 
Therefore, the monopole dominance in the string tension can be in principle investigated, as confirmed by numerical simulations on a lattice \cite{IKKMSS06}. 

Moreover, the existence of dimension--two condensate eliminates a tachyon mode  causing the Nielsen--Olesen instability of the vacuum with magnetic condensation.  Therefore, the restoration of the vacuum with magnetic condensation is obtained as a by-product of the above result.  
Furthermore, the existence of this condensate enables us to derive the Faddeev model describing glueballs as knot solitons as a low-energy effective theory of the Yang-Mills theory, as already pointed out in \cite{Kondo04,KOSSM06}. 
These are advantages of our reformulation of the Yang-Mills theory based on the non-linear change of variables.

\section*{Acknowledgments}
The author would like to thank Seikou Kato for carefully reading the manuscript and Toru Shinohara and Takeharu Murakami for some comments. 
This work is financially supported by Grant-in-Aid for Scientific Research (C) 18540251  from Japan Society for the Promotion of Science (JSPS), and in part by Grant-in-Aid for Scientific Research on Priority Areas (B)13135203 from the Ministry of Education, Culture, Sports, Science and Technology (MEXT).

\appendix
\section{New variables in the ortho-normal basis $(\bm{e}_1,\bm{e}_2,\bm{n})$}

\subsection{Covariant derivative in the ortho-normal basis}\label{app:cov-der}

For $\mathbb{V}_\mu= c_\mu \bm{n} + g^{-1} \partial_\mu \bm{n} \times \bm{n}$, we have  
\begin{align}
 \mathbb{V}_\mu \times \bm{e}_1 
&= c_\mu \bm{n} \times \bm{e}_1 + g^{-1} (\partial_\mu \bm{n} \times \bm{n}) \times \bm{e}_1
\nonumber\\
&= c_\mu \bm{e}_2 + g^{-1}[ (\partial_\mu \bm{n} \cdot \bm{e}_1) \bm{n}  - \partial_\mu \bm{n} (\bm{n} \cdot \bm{e}_1)]   
\nonumber\\
&= c_\mu \bm{e}_2 + g^{-1} (\partial_\mu \bm{n} \cdot \bm{e}_1) \bm{n}  .    
\end{align}
Then the covariant derivative of $\bm{e}_1$ in the background $\mathbb{V}_\mu$ reads 
\begin{align}
 D_\mu[\mathbb{V}] \bm{e}_1 
&:= \partial_\mu  \bm{e}_1 + g\mathbb{V}_\mu \times \bm{e}_1 
\nonumber\\
&= \partial_\mu  \bm{e}_1 + g c_\mu \bm{e}_2 +  (\partial_\mu \bm{n} \cdot \bm{e}_1) \bm{n} 
\nonumber\\
&= g c_\mu \bm{e}_2 + \partial_\mu  \bm{e}_1 -  (\bm{n} \cdot \partial_\mu \bm{e}_1) \bm{n}     .
\end{align}
Now we define 
$
 \bm{f}_\mu := \partial_\mu  \bm{e}_1 -  (\bm{n} \cdot \partial_\mu \bm{e}_1) \bm{n}   
$.
Then we can show easily that $\bm{f}_\mu$ is orthogonal to $\bm{e}_1$ and $\bm{n}$, and hence  $\bm{f}_\mu$ is proportional to $\bm{e}_2$.  Therefore  $\bm{f}_\mu$ is expressed as 
\begin{equation}
 \bm{f}_\mu = (\bm{f}_\mu \cdot \bm{e}_2) \bm{e}_2 
= (\bm{e}_2  \cdot \partial_\mu  \bm{e}_1) \bm{e}_2  
=: g h_\mu \bm{e}_2 , 
\quad
 h_\mu = g^{-1} (\bm{e}_2  \cdot \partial_\mu  \bm{e}_1) . 
\end{equation}
Thus we obtain
\begin{align}
 D_\mu[\mathbb{V}] \bm{e}_1 
= g c_\mu \bm{e}_2 + g h_\mu \bm{e}_2   
= gG_\mu  \bm{e}_2, \quad G_\mu = c_\mu  + h_\mu .
\label{DVe1}
\end{align}
In the similar way to the above, we can show that 
\begin{align}
 D_\mu[\mathbb{V}] \bm{e}_2 
= - gG_\mu  \bm{e}_1 .  
\label{DVe2}
\end{align}

If $\mathbb{X}_\mu$ is written in terms of  the orthonormal frame $(\bm{e}_1 , \bm{e}_2, \bm{n})$, 
\begin{align}
  \mathbb{X}_\mu = X_\mu^1 \bm{e}_1 + X_\mu^2 \bm{e}_2 ,
\end{align}
then we obtain 
\begin{align}
  D_\rho[\mathbb{V}] \mathbb{X}_\mu =& D_\rho[\mathbb{V}] (X_\mu^1 \bm{e}_1 + X_\mu^2 \bm{e}_2) 
\nonumber\\
=& X_\mu^1 D_\rho[\mathbb{V}] \bm{e}_1 + X_\mu^2 D_\rho[\mathbb{V}] \bm{e}_2 
+ \partial_\rho X_\mu^1 \bm{e}_1  + \partial_\rho X_\mu^2 \bm{e}_2  
\nonumber\\
=& X_\mu^1 g G_\rho \bm{e}_2 - X_\mu^2 g G_\rho \bm{e}_1 
+ \partial_\rho X_\mu^1 \bm{e}_1  + \partial_\rho X_\mu^2 \bm{e}_2  ,
\end{align}
and
\begin{align}
&  - D_\rho[\mathbb{V}]D_\rho[\mathbb{V}] \mathbb{X}_\mu 
\nonumber\\
=& 
 - X_\mu^1 g G_\rho D_\rho[\mathbb{V}] \bm{e}_2 + X_\mu^2 g G_\rho D_\rho[\mathbb{V}] \bm{e}_1 
\nonumber\\
&
- D_\rho[\mathbb{V}] [\partial_\rho X_\mu^1 \bm{e}_1]  - D_\rho[\mathbb{V}] [\partial_\rho X_\mu^2 \bm{e}_2]  
 - \partial_\rho[X_\mu^1 g G_\rho]  \bm{e}_2 + \partial_\rho [X_\mu^2 g G_\rho] \bm{e}_1 
\nonumber\\
=& 
 - X_\mu^1 g G_\rho D_\rho[\mathbb{V}] \bm{e}_2 + X_\mu^2 g G_\rho D_\rho[\mathbb{V}] \bm{e}_1 
\nonumber\\
&
- D_\rho[\mathbb{V}] [ \bm{e}_1]\partial_\rho X_\mu^1  - D_\rho[\mathbb{V}] [ \bm{e}_2]\partial_\rho X_\mu^2  
- \partial_\rho [\partial_\rho X_\mu^1] \bm{e}_1   - \partial_\rho [\partial_\rho X_\mu^2 ] \bm{e}_2   
 - \partial_\rho[X_\mu^1 g G_\rho]  \bm{e}_2 + \partial_\rho [X_\mu^2 g G_\rho] \bm{e}_1 
\nonumber\\
=& 
    g^2 G_\rho G_\rho X_\mu^1 \bm{e}_1 +  g^2 G_\rho G_\rho X_\mu^2 \bm{e}_2
\nonumber\\
&
- g G_\rho \bm{n}_2 \partial_\rho X_\mu^1  + g G_\rho \bm{n}_1 \partial_\rho X_\mu^2  
- \partial_\rho [\partial_\rho X_\mu^1] \bm{e}_1   - \partial_\rho [\partial_\rho X_\mu^2 ] \bm{e}_2   
 - \partial_\rho[X_\mu^1 g G_\rho]  \bm{e}_2 + \partial_\rho [X_\mu^2 g G_\rho] \bm{e}_1 
.
\end{align}
Thus we obtain
\begin{align}
&  \mathbb{X}_\mu \cdot [ - D_\rho[\mathbb{V}]D_\rho[\mathbb{V}]] \mathbb{X}_\mu 
\nonumber\\
=&    g^2 G_\rho G_\rho \mathbb{X}_\mu  \cdot \mathbb{X}_\mu 
+   (-X_\mu^1 \partial^2 X_\mu^1 - X_\mu^2 \partial^2 X_\mu^2 ) 
\nonumber\\
 &- X_\mu^2 gG_\rho \partial_\rho X_\mu^1  +  X_\mu^1 gG_\rho \partial_\rho X_\mu^2 
- X_\mu^2 \partial_\rho (gG_\rho X_\mu^1) + X_\mu^1 \partial_\rho (gG_\rho X_\mu^2)
\nonumber\\
=&    
  X_\mu^a [ - \partial^2 \delta^{ab} + g^2 G_\rho G_\rho  \delta^{ab} + 2g\epsilon^{ab}  G_\rho \partial_\rho + g \epsilon^{ab}   \partial_\rho G_\rho ] X_\mu^b 
\nonumber\\
=&    
  X_\mu^+ [ - \partial^2   + g^2 G_\rho G_\rho   + i( 2g G_\rho \partial_\rho + g   \partial_\rho G_\rho) ] X_\mu^-
\nonumber\\
&+ X_\mu^- [ - \partial^2   + g^2 G_\rho G_\rho  - i( 2g G_\rho \partial_\rho + g   \partial_\rho G_\rho) ] X_\mu^+ 
\nonumber\\
=& X_\mu^+ [ -(\partial_\rho-igG_\rho)^2  ] X_\mu^- + X_\mu^- [  -(\partial_\rho+igG_\rho)^2  ] X_\mu^+ .
\end{align}

\subsection{New variables and conventional Abelian projection}\label{app:transf}

For the orthonormal basis $(\bm{e}_1(x), \bm{e}_2(x), \bm{n}(x))$, we have
\begin{align}
  \mathbb{X}_\mu =& X_\mu^1 \bm{e}_1 + X_\mu^2 \bm{e}_2, 
\quad
 X_\mu^1 = \bm{e}_1 \cdot \mathbb{X}_\mu ,
\quad 
 X_\mu^2 =  \bm{e}_2 \cdot  \mathbb{X}_\mu,
\\ 
  \mathbb{B}_\mu =& B_\mu^1 \bm{e}_1 + B_\mu^2 \bm{e}_2,
\quad
 B_\mu^1 = g^{-1} \partial_\mu \bm{n} \cdot \bm{e}_2 ,
\quad 
 B_\mu^2 = - g^{-1} \partial_\mu \bm{n} \cdot \bm{e}_1 ,
\\
 h_\mu =&  g^{-1}   \partial_\mu \bm{e}_1  \cdot  \bm{e}_2
= - g^{-1} \partial_\mu \bm{e}_2  \cdot \bm{e}_1      .
\end{align}
For this basis, two gauge transformations are expressed as follows. 

Gauge transformation I: $\bm{\omega}(x) = \omega_1(x) \bm{e}_1(x) + \omega_2(x) \bm{e}_2(x) + \theta(x) \bm{n}(x)$
\begin{align}
  \delta_\omega c_\mu =& \partial_\mu \theta + ig(X_\mu^{\pm} \omega^{\mp} - X_\mu^{\mp} \omega^{\pm}) ,
\\
  \delta_\omega X_\mu^{\pm} =& [ \partial_\mu + ig (c_\mu+h_\mu)] \omega^{\pm} \mp ig X_\mu^{\pm} \theta ,
\\
 \delta_\omega h_\mu  =& 0 ,
\end{align}
Gauge transformation II: $\bm{\omega'}(x) = \omega_1'(x) \bm{e}_1(x) + \omega_2'(x) \bm{e}_2(x) + \theta'(x) \bm{n}(x)$
\begin{align}
  \delta_{\omega'} c_\mu  =& \partial_\mu \theta' + ig(B_\mu^{\pm} \omega'{}^{\mp} - B_\mu^{\mp} \omega'{}^{\pm}) ,
\\
  \delta_{\omega'} X_\mu^{\pm} =& 0 ,
\\
 \delta_{\omega'} h_\mu  =& -  \delta_{\omega'} c_\mu  ,
\end{align}
where we have defined
$
 \mathcal{O}^{\pm} := \frac{1}{\sqrt{2}} (\mathcal{O}^{1}+ i \mathcal{O}^{2}) . 
$

In this basis, the nMAG reduces to the conventional MAG apparently: 
\begin{equation}
 D_\mu[\mathbb{V}]\mathbb{X}_\mu =0 \Longleftrightarrow 
 \partial_\mu X_\mu^a -g(c_\mu+h_\mu) \epsilon^{ab}X_\mu^b=0 ,
\end{equation}
provided that the Abelian part $a_\mu$ in the conventional Abelian projection is identified as
\begin{equation}
 a_\mu \leftrightarrow c_\mu+h_\mu := G_\mu , 
\quad 
 A_\mu^a \leftrightarrow  X_\mu^a ,
\end{equation}
since 
$
 D_\mu[\mathbb{V}]\mathbb{X}_\nu = \bm{e}_1 [\partial_\mu X_\nu^1 -g(c_\mu+h_\mu)X_\nu^2] + \bm{e}_2 [\partial_\mu X_\nu^2 +g(c_\mu+h_\mu)X_\nu^1]
$.
It should be remarked that $a_\mu$ is invariant under the SU(2) gauge transformation II:
\begin{align}
  \delta_{\omega'} G_\mu = 0 . 
\end{align}
This is also the case for $X_\mu^a$:
\begin{align}
 \delta_{\omega'} X_\mu^a = 0. 
\end{align}
This shows that this is different from the naive Abelian projection.

\subsection{Jacobian for the non-linear change of variables}\label{app:Jacobian}

The Jacobian is calculated as follows.%
\footnote{
The author is indebted to Toru Shinohara for completing this part of Appendix. 
}
By using the above bases, the field $\mathscr{A}_\mu^A$ is decomposed as
\begin{align}
  \mathscr{A}_\mu = c_\mu \bm{n} + \mathbb{B}_\mu + \mathbb{X}_\mu 
= c_\mu \bm{n} + (B_\mu^a + X_\mu^a ) \bm{e}_a , \quad
\mathscr{A}_\mu^A = c_\mu n^A + (B_\mu^a + X_\mu^a ) e_a^A.
\label{decomp}
\end{align}
We consider the change of $3D+2$ variables:
\begin{align}
 (\mathscr{A}_\mu^A,n^B) \rightarrow (c_\nu, X_\nu^a,n^C) \quad (a=1,2; A,B,C=1,2,3; \mu,\nu=0, \cdots, D-1) . 
\end{align}
Here $n^B$ and $n^C$ should be understood as denoting two independent degrees of freedom obtained after solving the constraint $n^A n^A=1$. However, if we choose specific components (directions), the color symmetry is apparently broken. Therefore, we keep this notation in the followings, keeping this convention in mind. 

The Jacobian is the determinant for the $(3D+2)\times(D+2D+2)$ matrix:
\begin{align}
 d\mathscr{A}_\mu^A dn^B  = J dc_\nu dX_\nu^a  dn^C , 
\quad
 J = 
 \begin{vmatrix}
  \frac{\partial \mathscr{A}_\mu^A}{\partial c_\nu} & \frac{\partial \mathscr{A}_\mu^A}{\partial X_\nu^a} & \frac{\partial \mathscr{A}_\mu^A}{\partial n^C} 
  \cr
  \frac{\partial n^B}{\partial c_\nu} & \frac{\partial n^B}{\partial X_\nu^a} & \frac{\partial n^B}{\partial n^C} \cr
 \end{vmatrix} .
\end{align}
Since $c_\nu, X_\nu^b, n^C$ are independent, we have
\begin{align}
 \frac{\partial n^B}{\partial c_\nu} = 0 , \quad
\frac{\partial n^B}{\partial X_\nu^a} = 0 , \quad
 \frac{\partial n^B}{\partial n^C} = \delta^{BC} .
\end{align}
Then the Jacobian reduces to the determinant for the $3D\times(D+2D)$ matrix:
\begin{align}
 J = 
 \begin{vmatrix}
  \frac{\partial \mathscr{A}_\mu^A}{\partial c_\nu} & \frac{\partial \mathscr{A}_\mu^A}{\partial X_\nu^a} & \frac{\partial \mathscr{A}_\mu^A}{\partial n^C} \cr
 0  & 0 & 1 \cr
 \end{vmatrix} 
=  \begin{vmatrix}
   \frac{\partial \mathscr{A}_\mu^A}{\partial c_\nu} & \frac{\partial \mathscr{A}_\mu^A}{\partial X_\nu^a} \cr
 \end{vmatrix} .
\end{align}
Making use of (\ref{decomp}), we have
\begin{align}
 \frac{\partial \mathscr{A}_\mu^A}{\partial c_\nu} = \delta_{\mu\nu} n^A , \quad
  \frac{\partial \mathscr{A}_\mu^A}{\partial X_\nu^a} = \delta_{\mu\nu} e_a^A ,
\end{align}
and we conclude
\begin{align}
 J = 
  \begin{vmatrix}
    \delta_{\mu\nu} n^A  & \delta_{\mu\nu} e_a^A \cr
 \end{vmatrix}
=    \begin{vmatrix}
      n^A  &  e_a^A \cr
 \end{vmatrix}
= | \bm{n} \bm{e}_1 \bm{e}_2 |
= | \bm{n} \cdot (\bm{e}_1 \times \bm{e}_2) |
= 1  .
\end{align}


\end{document}